%% file: MyFrLe-2014-10-29.tex
\documentclass[preprint]{elsarticle}   	
\usepackage{geometry}                		
\usepackage[parfill]{parskip}    		
\usepackage{graphicx}				
\usepackage{amsmath,amssymb,amsfonts}
\usepackage{algorithm2e}

\usepackage[usenames,dvipsnames]{color}

\input{macros.tex}

\fntext[fn1]{This author was supported by a grant as part of research programme 613.001.009, which is financed by the Netherlands Organisation for Scientific Research (NWO).}
\fntext[fn2]{This author was supported by grant EPSRC EP/G036136/1.}
\cortext[cor1]{Corresponding author}

\address[cwi]{Centrum Wiskunde \& Informatica, P.O. Box 94079, 1090 GB Amsterdam, the Netherlands}
\address[ed]{School of Mathematics and Maxwell Institute for Mathematical Sciences, University of Edinburgh, James Clerk Maxwell Building, Kings Buildings Edinburgh EH9 3JZ}
\address[uu]{Mathematical Institute, Utrecht University, P.O. Box 80010, 3508 TA Utrecht, the Netherlands}

\usepackage{caption}
\usepackage{subcaption}

\newcommand{\figwidth}{.35}

\begin{document}

\title{Least-biased correction of extended dynamical systems using observational data}
\author[cwi]{Keith W. Myerscough\corref{cor1}\fnref{fn1}}
\ead{k.w.myerscough@cwi.nl}

\author[uu]{Jason Frank}
\ead{j.e.frank@uu.nl}

\author[ed]{Benedict Leimkuhler\fnref{fn2}}
\ead{b.leimkuhler@ed.ac.uk}

\begin{abstract}
We consider dynamical systems evolving near an  equilibrium statistical state where the interest is in modelling long term behavior that is consistent with thermodynamic constraints.   We adjust the distribution using an entropy-optimizing formulation that can be computed on-the-fly, making possible partial corrections using incomplete information,  for example measured data or data computed from a different model (or the same model at a different scale).    We employ a thermostatting technique to sample the target distribution with the aim of capturing relavant statistical features while introducing mild dynamical perturbation (thermostats).  The method is tested for a point vortex fluid model on the sphere, and we demonstrate both convergence of equilibrium quantities and the ability of the formulation to balance stationary and transient-regime errors.    
\end{abstract}


\maketitle


\section{Introduction}
In many applications of modern computational science the physical laws (and equations of motion) are well established yet the detailed behavior is unpredictable on long time scales due to the presence of deterministic chaos.   Examples of this arise in molecular dynamics modelling \cite{md1,md2} and in the study of turbulent fluids in the atmosphere and ocean \cite{HoLuBe:98,Davidson:04}.  For these problems, long simulations are routinely run, despite the  lack of predictability, in the hope that the resulting simulation will yield useful statistical knowledge (e.g. the statistics of rare transitions between basins in molecular dynamics, or slow relaxation processes in fluids).  
We refer to this approach  as \emph{dynamical sampling}, where the name is suggestive of the typical  requirement that simulated paths are sufficiently accurate to allow the computation of measures of dynamical mixing such as two-point temporal correlation functions \cite{shadow1}.   

In Hamiltonian systems such as molecular dynamics, it is common to run canonically prepared ensembles of microcanonical (i.e.~constant energy) simulations in order to minimize the perturbation of dynamical properties.  For such systems, backward error analysis \cite{Hairer:06,Leimkuhler:04} suggests that the global behavior can best be understood not as 
the approximation of particular trajectory but rather as an accurate path for a perturbed continuum process described by modified equations.  In the case of dynamic sampling of complex systems,  the statistics of simulation data are therefore biased in that they sample an invariant measure of the modified equations, i.e. bias arises as an artifact of time discretization. Statistical bias may also arise due to spatial discretization.    For example, in the setting of geophysical fluid dynamics, a comparison of discretizations of the quasi-geostrophic equations reveals that the long time mean potential vorticity field and pointwise fluctuation statistics are heavily dependent on discrete conservation laws such as energy, enstrophy, and material conservation of vorticity \cite{AbMa03,DuFr07,DuFr10}.  It is usually impossible to construct numerical discretizations that automatically preserve all conservation laws of statistical relevance for a given problem, so the discretization necessarily perturbs the 
statistical distribution.    The discretization bias may be 
reduced by refining the discretization or by incorporating a Metropolis condition \cite{MeRoRoTeTe:53}, but such techniques also typically lead to an increase in computational overhead, which may be unacceptable in large scale applications.

The combination of the need for computations to address both the stationary constraint (``nearness to the steady-state distribution'') and to provide accuracy with respect to dynamical processes poses difficult challenges for the simulator.    In this paper we consider an approach to perturbing dynamics to correct statistical bias in systems at statistical equilibrium.    If the  statistical distribution is completely specified via a probability density function (pdf) it can be sampled using a ``thermostat.''  Such thermostats, originating in molecular dynamics, can be extended to handle both smooth \cite{BaLaLe:03} and nonsmooth \cite{BaFrLe:13} densities and to treat noncanonical Hamiltonian systems.  In \cite{Dubinkina:10}, a thermostat was used as a model reduction technique for a vortex model of a fluid (suppressing the detailed interactions of a few strong vortices with a weak vortex field). In another recent article \cite{DiFaBaChSkNe:pre}, thermostats have been suggested as a means of  sampling 
incompletely specified systems (with noisy gradients), with applications in learning theory.  The standard framework of thermostating used in these and other applications assumes a fixed, known distribution such as the Gibbs-Boltzmann distribution.   In this article, we assume that, instead of the pdf, what is available is a partial set of  \emph{expectations} of observables 
with 
respect to the unknown invariant measure, which may arise from experiment or other types of modelling.  In this setting, information theory (in particular entropy maximization \cite{Jaynes57a,Jaynes57b}) offers tools for constructing least-biased densities,  close to some known prior distribution, which are consistent with observations. The iterative method (based on \cite{AgAlLe79,Haken06,DaGu12}) involves computation of Lagrange multipliers (one for each observable) that modify the probability density.  The Lagrange multipliers are computed using an iterative procedure in which each stage represents an ensemble average (with respect to the previous  estimate of the density).  To make the method practical in situations where the sampling is costly, we consider an adaptive procedure which uses only short-time ensemble bursts to gradually tune the parameters in simulation.       At the same time, we are able to show in numerical experiments that autocorrelation functions are only modestly perturbed meaning 
that we would expect to be able to recover dynamical information such as diffusion and other transport coefficients.     

We emphasize that the framework of least-biased estimation is well known but applied here in a novel way.  A related technique is used by Majda and Gershgorin \cite{MaGe11} to develop a framework for validating computational models and choosing the optimal linear combination of an ensemble of model outputs, so as to minimize the discrepancy between the ensemble distribution and the least-biased estimate, arguing that the latter is the best available measure for comparison, when the true invariant measure is unknown.  With the approach we develop here, we enforce exact adherence to the least-biased measure, which is constructed automatically in simulation, by perturbing the dynamics to take full advantage of available information.  

The remainder of this article is organized as follows.  In the next section, we discuss the maximum entropy framework for correcting the density to reflect thermodynamic constraints.   We apply and evaluate the method in the setting of a system of point vortices on the surface of a sphere, which represents a simple geophysical model with multiple statistically relevant first integrals.



\section{Bias correction method}
\label{sec:Method}
Our interest is in extended dynamical systems with many degrees of freedom that evolve near statistical equilibrium.  Further, we imagine that we are given a simplified dynamical model for the evolution of some projection (i.e. a ``coarse graining'') of the phase variables (coarse grained variables $y(t)\in \R^d$) .   Although the original system is complex and its details unknown, we assume that we can obtain in some way (e.g. through measurement) a collection of ``observations''  of mean values of functions of the reduced variables. That is there are functions $C_k:\R^d\rightarrow \R$, $k=1,2\ldots,K$ and given values $c_k$, $k=1,2,\ldots, K$, such that
\begin{equation}\label{obs}
	c_k =\langle C_k(y) \rangle, \quad k=1,\dots,K,
\end{equation}
where $\langle C_k(y) \rangle$ represents averaging with respect to the true, empirical invariant measure of the dynamical system.
Our goal is to find a perturbed dynamical model for the reduced variables which (a)  is compatible with the indicated thermodynamic constraints (\ref{obs}), and (b) weakly perturbs the dynamics compared to those of the native model.

%

%
%
Empirical information theory generalizes the principle of insufficient reason, by proposing the least-biased probability density consistent with a set of observations.  See the classical work of Jaynes \cite{Jaynes57a,Jaynes57b}, the monographs \cite{Haken06,Dewar14} and an extensive treatment in the geophysical fluid context in the monograph by Majda and Wang \cite{Majda:06}.  The least-biased density is defined as the probability density $\rho(y)$ that maximizes the information entropy 
functional 
\[
	\mathcal{S} [\rho] = - \int_\mathcal{D} \rho(y) \log \rho(y) \, dy,
\]
subject to a set of constraints given by observations.
When $\mathcal{D}$ is a compact set and there are no observations, the minimizer is the uniform density $\rho \equiv |\mathcal{D}|^{-1}$.   The entropy $\mathcal{S}$ is the unique measure of uncertainty that is positive valued, monotonically increasing as a function of uncertainty, and additive for independent random variables. With observable functions $\{ C_k(y) | k=1,2,\ldots K\}$ let
\begin{equation}\label{expect}
	\mathbb{E}_\rho C_k = \int_\mathcal{D} C_k(y) \rho(y) \, dy
\end{equation}
denote expectation in the (as yet undetermined) density $\rho$. 
Defining Lagrange multipliers $\lambda_k$, $k=1,\dots, K$, associated with the observables $C_k$, the constrained minimization problem is 
\[
	\rho = \arg \max_{\hat{\rho}} \left[ \mathcal{S}[\hat{\rho}] - \sum_{k=1}^K \lambda_k \left( \mathbb{E}_{\hat{\rho}} C_k(y)  - c_k \right) \right].
\]
When it exists, the maximum entropy solution satisfies
\[
	\rho(y) = \lambda_0 \exp \left( -\lambda_1 C_1(y) - \dots -\lambda_K C_K(y) \right),
\]
where $\lambda_0$ is chosen to satisfy $\int_\mathcal{D} \rho \, dy = 1$, and $\lambda_k$ is chosen such that $\mathbb{E}_\rho C_k(y) = c_k$.


In some cases, besides the observations, we may be given prior statistical information on the process $y(t)$.
The Kullback-Leibler divergence, or relative entropy,
\[
	\mathcal{S}[\rho(y)] = \int \rho(y) \ln \frac{\rho(y)}{\pi(y)} \, dy
\]
which represents a (non-symmetric) distance between measures.  It quantifies the information lost in approximating $\rho(y)$ by $\pi(y)$.  

Suppose $y$ is a random variable with distribution (law) $y\sim\rho$, where $\rho$ is unknown.
Suppose further, that we are given a prior distribution $\pi$, presumed to be close to $\rho$, and a set of $K$ observations (\ref{obs}).
Following Jaynes  \cite{Jaynes57a,Jaynes57b}, the least-biased distribution $\rho$ consistent with the observations $c_k$ and prior $\pi$ solves the constrained minimization problem
\[
	\rho = \arg\min_\rho \left[ \mathcal{S}  - \lambda_0 \left(1 - \int \rho(y) \, dy\right)- \sum_{k=0}^K \lambda_k \left(c_k - \int C_k(y) \rho(y) \, dy\right) \right],
\]
where the $\lambda_k$ are Lagrange multipliers to enforce the condition that the expectations (\ref{expect}) 
agree with the observations (\ref{obs}).  
The solution to the variational problem is
\begin{equation} \label{posterior} 
	\rho(y) = \lambda_0 \exp \left( - \lambda_1 C_1(y) - \dots - \lambda_K C_K(y) \right) \pi(y),
\end{equation}
where the Lagrange multipliers $\lambda_k$ are chosen consistently with the observations (\ref{obs}) and $\lambda_0$ is a normalization constant so that $\rho$ is a probability density function.

Methods for determining the Lagrange multipliers are discussed in \cite{AgAlLe79,Haken06,DaGu12}.
We use the following algorithm based on re-weighting.  
Assume we are given a sequence of samples $y^n$, $n=1, \dots, N$, distributed according to a known prior distribution $\pi(y)$, i.e.~$y^n \sim \pi$.  
The expectation under $\pi(y)$ of a function $\Phi(y)$ has the consistent and unbiased estimator
\[
	\widehat\Phi^\pi = \frac{1}{N} \sum_{n=1}^N \Phi(y^n).
\]

Given the posterior distribution $\rho(y)$ of the form (\ref{posterior}), compute the expectation $\mathbb{E}_\rho \Phi$ by re-weighting of the integral
\[
	\mathbb{E}_\rho \Phi = \int \Phi(y) \rho(y) \, dy =\lambda_0 \int \Phi(y) e^{-\sum_{i=1}^K \lambda_i C_i(y)} \pi(y)\, dy
	=\lambda_0 \mathbb{E}_\pi \{ \Phi(y) \lambda_0 e^{-\sum_{i=1}^K \lambda_i C_i(y)} \},
\]
yielding an unbiased estimator for $\mathbb{E}_\rho \Phi$ given by
\[
	\widehat\Phi^\rho = \frac{\lambda_0}{N} \sum_{n=1}^N \Phi(y^n) e^{-\sum_{i=1}^K \lambda_i C_i(y^n)}.
\]

We wish to ensure that the observations $c_k$ satisfy
\[
	c_k = \widehat C_k^\rho = \frac{\lambda_0}{N} \sum_{n=1}^N C_k(y^n) e^{-\sum_{i=1}^K \lambda_i C_i(y^n)}, \quad k=1,\dots,K.
\]
We can use this fact to define a Newton-Raphson iteration to determine the Lagrange multipliers $\lambda_k$.  Define the residual $r$ with components
\[
	r_k (\lambda) = c_k - \frac{\lambda_0}{N} \sum_{n=1}^N C_k(y^n) e^{-\sum_{i=1}^K \lambda_i C_i(y^n)}, \quad k=1,\dots,K,
\]
with $\lambda = (\lambda_1,\dots,\lambda_K)$ and $r = (r_1(\lambda),\dots,r_K(\lambda))$.  Note that $\lambda_0$ can be viewed as a function of $\lambda_1,\lambda_2,\ldots, \lambda_K$ chosen from the normalization condition, i.e.,
\[
\lambda_0 = \left [ \sum_{n=1}^N  e^{-\sum_{j=1}^K \lambda_j C_j(y^n)}\right ]^{-1}.
\]

The Jacobian matrix $J=(J_{kj})$ of the vector function $r$ is determined as
\[
	J_{kj}(\lambda) := \frac{\partial r_k}{\partial \lambda_j} = \frac{\lambda_0}{N} \sum_{n=1}^N  C_k(y^n) C_j(y^n) e^{-\sum_{i=1}^K \lambda_i C_i(y^n)} \quad j,k=1,\dots,K.
\]
The iteration then proceeds as $\lambda^{\alpha+1} \leftarrow \lambda^{\alpha} - J^{-1}(\lambda^{\alpha}) r(\lambda^{\alpha})$.

\subsection{Adaptive determination of Lagrange multipliers\label{sec:adaptive}}
In many cases it will be difficult or costly to carry out a complete sampling of the distribution at each iteration step of the Newton procedure.   Moreover, the standard framework excludes applications where (i) the statistical knowledge is expected to improve as the simulation progresses, (ii) the average observables are known to vary slowly with time, or (iii) it is unfeasible to constuct a large enough ensemble distributed in the prior. For these cases we consider using the simulation data of a small ensemble (propagated in short bursts of $M$ timesteps) for updating the Lagrange multipliers for mean observation data. This results in an adaptive algorithm for obtaining the Lagrange multipliers ``on-the-fly'' during simulation.  

Consider the following: an ensemble of $P$ simulations (preferably with initial conditions distributed close to $\pi(y)$) is advected $M \Delta t$ in time, where $M$ is chosen sufficiently large such that the ensemble members sample $\pi$ well. These ensemble members can be used in an estimator for $\mathbb{E}_{\lambda^1}C_k$ given by
\begin{equation*}
\widehat C_k^{(1)} = \frac{1}{P} \sum_{p=1}^P C_k(y_0^p) \lambda^1_0 \exp\left(-\sum_{i=1}^K \lambda^1_i C_i(y_0^p) \right), \quad k=1,\dots,K,
\end{equation*}
where the superscript $(1)$ indicates that it is an estimator for a distribution with Lagrange multipliers $\lambda^1_j$.
A Newton-Raphson iteration to find the first set of Lagrange multipliers such that observations match data has the residual
\begin{equation*}
r^1_k = \widehat C_k^{(1)} - c_k = \frac{1}{P} \sum_{p=1}^P C_k(y_0^p)  \lambda^1_0\exp\left(-\sum_{i=1}^K \lambda^1_i C_i(y_0^p) \right) - c_k, \quad k=1,\dots,K.
\end{equation*}

Using this updated value for $\lambda$ the simulations will sample the distribution $\rho \propto e^{-\sum_{i=1}^K \lambda^1_i c_i(y)}$ after some time $M\Delta t$.  (See below for some practical issues associated to this.) Using these samples alongside the initial data, $\lambda^2_j$ is found. Iteration of this process leads to the following equation
\begin{equation}
r^m_k = \widehat C_k^{(m)} - c_k = \frac{1}{mP} \sum_{p=1}^P  \sum_{\ell=0}^{m-1} C_k(y_{\ell M}^p) \lambda^{\ell}_0 \exp\left(\sum_{i=1}^K (\lambda^l_i-\lambda^m_i) C_j(y_{\ell M}^p) \right) - c_k, \; k=1,\dots,K,
\label{eq:update}
\end{equation}
where we remind that, at each stage of iteration, $\lambda^l_0$ is a function of the multiplier vector $\lambda^l$ (indices $1\ldots K$).
In the calculation (\ref{eq:update}) $\lambda^0_1,\ldots \lambda^0_K$ would ideally be zero. There are cases where it is impossible to obtain an accurate initial distribution according to the prior, in which case the initial Lagrange multipliers can be chosen different from zero if initial conditions sampling $\pi\lambda_0^0\exp\left(-\sum_k \lambda_k^0 C_k\right)$ are easier to find than those sampling just the prior $\pi$.
Solutions of \eqref{eq:update} are found using Newton-Raphson iteration. The gradient is given by
\begin{equation}
\frac{\partial r^m_k}{\partial \lambda^m_j} = \frac{1}{mP} \sum_{p=1}^P \sum_{\ell=0}^{m-1} C_k(y_{\ell M}^p) C_j(y_{\ell M}^p) \lambda^{\ell}_0 \exp\left(\sum_{i=1}^K (\lambda^l_i-\lambda^m_i) C_i(y_{\ell M}^p) \right), \quad j,k=1,\dots,K.
\label{eq:updategrad}
\end{equation}
In this way the Lagrange multipliers may be found ``on-the-fly.''

As a convergence result let us consider the case where both $P$ and $M$ may be chose arbitrarily large. For $P \rightarrow \infty$ the Lagrange multipliers computing using only the initial data sampling the prior will be correct. Given sufficiently large $M$ the samples after evolving the thermsotated system $M\delta t$ in time will accurately sample the distribution corresponding to these Lagrange multipliers. The ensemble averages will then correspond to the observations, and the Lagrange multipliers no longer need updating.

\subsubsection{Adaptive algorithm\label{sec:practical}}
There are two important practical modifications to the algorithm described above that are included in the numerical implementation of this method:
\begin{itemize}
\item The first modification is limiting the rate of change of the Lagrange multipliers. If the Lagrange multipliers change rapidly, the thermostat may require a long time to equilibriate. This requires a larger value for $M$, increasing the simulation time required before including new samples. The effect is especially noticable at the beginning of a simulation, due to two factors: (i) the small sample size leads to inaccurate expectations for the observables, and (ii)  the Lagrange multipliers may be far from their correct value.  By limiting the rate of change of the Lagrange multipliers, these problems are circumvented.
\item The second modification regards the number of samples included when updating the Lagrange multipliers. In equations \eqref{eq:update} and \eqref{eq:updategrad} \emph{all} previous values $\lambda_k^{\ell}$ are included. In a long simulation, this leads to a growing computational demand. By taking only a fixed number ($q$) of recent steps the computational demand can be reduced. In the case that the initial samples cannot accurately be drawn from the prior, this has the further advantage that these inaccuracies are eventually forgotten.
\end{itemize}

The algorithm, including these practical modifications, is summarized in Algorithm \ref{alg:lm}.
\begin{center}
\begin{algorithm}[H]
\caption{Adaptive determination of Lagrange multipliers ``on-the-fly''}
\label{alg:lm}
Given initial conditions according to prior $\pi(y)$\\
Set initial Lagrange multipliers to zero\\
\For{$m\leftarrow 1$ \KwTo $n$}{ 
  \For{$j\leftarrow 1$ \KwTo $M$}{ 
    advance simulation one time step using current value for the Lagrange multipliers
  }
  store relevant simulation observables for time step $im$.\\
  \While{$|\ensbra{C_k(y)}_\lambda - c_k| > \mathrm{tolerance}$ }{
    compute residual using reweighted samples at times $M \times {\rm max}(m-q,0),\ldots,mM$\\
    compute residual gradient using the same data\\
    update Lagrange multiplier estimation
  }
  limit the change in Lagrange muliplier (if necessary)
}
\end{algorithm}
\end{center}

\subsection{Thermostat}
\label{sec:td}
We introduce the bias-correction methodology for a Hamiltonian dynamical system
\begin{equation}\label{HODE}
\frac{dy}{dt} = f(y) = B(y) \nabla H(y), \qquad y(t)\in\mathcal{D},~B(y)=-B(y)^T,~H(y):\mathcal{D}\rightarrow \R,
\end{equation}
possessing a divergence-free vector field\footnote{The latter condition is automatic for systems (\ref{HODE}) with constant $B$. Strictly speaking, the approach described here is applicable to any system with divergence-free vector field $\nabla\cdot f \equiv 0$ possessing one or more first integrals.} $\nabla\cdot f\equiv 0$.
Invariance of the Hamiltonian $H$ along solutions of (\ref{HODE}) follows from $\frac{d}{dt} H(y(t))) = \nabla H \cdot \frac{dy}{dt} = \nabla H \cdot B \nabla H = 0$, due to skew-symmetry of $B(y)$.  Additional first integrals are often present:  $I_\ell(y): \nabla I_\ell \cdot f \equiv 0$, $\ell=1,\dots,L$.  In this paper we consider only the case where all observables of the physical process $\mathcal{Y}(t)$ correspond to functions of the conserved quantities $\{ H, I_\ell, \ell=1, \dots,L\}$, that is, $C_k(y) = C_k(H(y), I_1(y),\dots,I_L(y))$, $k=1,\dots,K$.

Thermostats are used in molecular dynamics to model the trajectories of molecules in a fluid at constant temperature.  From statistical mechanics, it is well known that the trajectories of a system of particles in thermal equilibrium with a reservoir at constant temperature sample the canonical or Gibbs distribution, which has global support.  The governing equations are Hamiltonian, however, implying that the trajectories are restricted to a level set of Lebesgue measure zero.  Hence, to model a system at constant temperature, it is necessary to perturb the vector field to make trajectories ergodic with respect to the Gibbs distribution.   The most common way of achieving this is by adding suitable stochastic and dissipative terms satisfying a fluctuation-dissipation relation (Langevin dynamics).  An advantage of Langevin dynamics is provable ergodicity with respect to the Gibbs distribution \cite{Mattingly:02}.  However, DelSole \cite{DelSole00} warns that direct stochastic forcing of trajectories leads to 
inaccurate dynamical quantities since autocorrelation functions are strongly perturbed.  For smooth deterministic Hamiltonian dynamics, normalized velocity autocorrelation functions are of the form $1 - c \tau^2$, $c>0$ in the zero-lag limit $\tau\rightarrow 0$, whereas the autocorrelation of a variable that is directly forced by white noise must take the form $\exp(-\kappa\tau)$, $\kappa>0$ in the same limit.  This implies that the direct stochastic perturbation leads to auto-correlation functions that have nonzero slope and opposite curvature at zero lag.

An alternative approach, pioneered by Nos\'{e} \cite{Nose84a,Nose84b} and Hoover \cite{Hoover85} proceeds to augment the phase space by one dimension through coupling of (\ref{HODE}) to an additional thermostat variable $\xi(t)$.  The dynamics of $\xi$ are constructed to ensure that the extended dynamics on $\R^{d+1}$ preserves an equilibrium density whose marginal on $\R^d$ is the target (e.g.~Gibbs) density.  The fully deterministic thermostats of Nos\'{e} and Hoover have no mechanism to guarantee ergodicity with respect to the target density, and hence have been modified by various authors who include stochastic forcing of the thermostat variable $\xi$, leading to the so-called Nos\'{e}-Hoover-Langevin method \cite{SaDeCh07,Leimkuhler:09,Leimkuhler:11}.  A generalization to generic Hamiltonian systems is the Generalized Bulgac-Kusnezov (GBK) \citep{Leimkuhler:10,Bulgac:90} thermostat: 
\begin{subequations}
\label{eq:GBK}
\begin{align}
  dy &= f(y) dt + \xi \epsilon g(y) dt \label{eq:GBK1}\\
  d\xi    &= \epsilon h(y) dt - \gamma \xi dt + \sqrt{2 \gamma} dw, 
  \label{eq:GBK2}
\end{align}
\end{subequations}
where $\varepsilon>0$ and $\gamma>0$ are parameters, $w(t)$ is a scalar Wiener process, and $g$ and $h$ are discussed below.  Given a target density $\rho(y) \propto \exp(-A(y))$, $A:\mathcal{D}\rightarrow \R$, denote the augmented product density by $\tilde{\rho}(y,\xi) = \rho(y)\cdot \mu(\xi)$, with $\mu$ a univariate normal distribution with mean zero and standard deviation one.  It is easily checked that $\tilde{\rho}$ is stationary under the Fokker-Planck operator associated with (\ref{eq:GBK}) provided
\begin{equation}\label{eq:hfun}
	h(y) = \nabla \cdot g - g \cdot \nabla A.
\end{equation}
Furthermore it is argued in \cite{BaFrLe:13} that the target measure is ergodic provided the vector fields $f$ and $g$ satisfy a H\"{o}rmander condition.  In some cases it is also desirable to use the freedom in choosing $g$ to ensure preservation of some first integrals of the vector field $f$.  We will see an example of this later in this paper.

The parameter $\varepsilon$ can be used to control the relative strength of the thermostat compared to that of the unperturbed vector field $f$.  This will affect the rate at which the invariant measure is sampled, but has no influence on the measure itself.  It has been proved in \cite{Leimkuhler:11} and observed numerically in \cite{BaFrLe:11,BaFrLe:13} that GBK/NHL thermostating leads to a weak perturbation of the original trajectories in the sense that autocorrelation functions preserve the leading terms, i.e. have the form $1 - c\tau^2+O(\tau^3)$, as $\tau\rightarrow 0$.  The GBK thermostat is applicable when the vector field $f$ is divergence free $\nabla\cdot f \equiv 0$ and when the target density $\rho$ is a function of first integrals of $f$.   

We therefore propose (1) constructing a least-biased information theoretic target density based on observations of functions of conserved quantities (with or without prior distribution), followed by  (2) thermostated perturbation of dynamics to ensure sampling of the target distribution with a GBK thermostat.  The thermostating method is incorporated into Algorithm 1 to provide the scheme for sampling the adapted, data-dependent distribution.

\section{Application to reduced modelling of point vortices}

In this section we apply the least-biased correction methodology to the simple model of point vortices on the sphere.  We choose this model because it has a Poisson structure and multiple conserved quantities, including total energy and angular momentum and a set of Casimirs, with various degrees of statistical significance.  Although one can construct a point vortex approximation of quasi-geostrophic potential vorticity dynamics, we ignore the effects of topography and finite deformation radius for computational simplicity.

\subsection{Point vortex system}

A simple conceptual model of the atmosphere is given by the quasigeostrophic potential vorticity equation on a rotating sphere:
\begin{equation}\label{eq:QG}
	\frac{Dq}{Dt} \equiv \frac{\partial q}{\partial t} + u\cdot \mathrm{grad}\, q = 0, \quad 
	\mathrm{div}\, u = 0, 
	\quad q = \mathrm{curl}\, u + f_0 + h,
\end{equation}
where $q$ denotes the potential vorticity, $u$ is the velocity field in the tangent plane, assumed divergence-free, $f_0 = 2\Omega \sin \theta$ is the local Coriolis force, and $h$ is the surface topography.  

The vortex approximation of (\ref{eq:QG}) is well known.  For algorithms and analysis of the dynamics of point vortex systems, see the books \cite{Newton:01,Majda:02}.  For advanced modelling and convergence analysis in the continuum limit, see \cite{Cottet:00}.
For numerical computation with point vortices, it is advantageous to embed the sphere in $\R^3$.  In the sequel we will denote vectors in $\R^3$ by bold type.  For simplicity we neglect topography, taking $h\equiv 0$, under which assumption  the quasigeostrophic model is equivalent to the 2D Euler equations.  
We may then also ignore rotation (i.e.~$f_0\equiv 0$) as it gives rise to a trivial rigid body rotation of the ensuing point vortex system \citep{Newton:01}.
Since the velocity field is divergence-free in the tangent plane, it can be represented in terms of a stream function $\psi$ as
\[
	\bs{u} = \hat{\bs{k}} \times \nabla \psi
\]
where $\bs{\hat{k}}$ is the unit normal vector on the surface of the sphere.  
The potential vorticity and stream function are related by $\Delta \psi = q - f_0 - h$ with $\Delta$ the Laplace-Beltrami operator (from which it is apparent that topography, if included, would lead to a nonhomogeneous background term in the stream function).

A point vortex system is constructed by taking the vorticity field in (\ref{eq:QG}) to be a sum of Dirac distributions
\[
	q(x,t) = \sum_{i=1}^M  \Gamma_i \delta (\bs{x} - \bs{x}_i),
\]
where $\Gamma_i$ is the vortex strength or circulation of the $i$th point vortex. 
The point vortices induce a stream function 
$\psi(\bs{x}) = \sum_i \frac{-1}{4\pi} \sum \Gamma_i \ln \left( 2 - 2\bs{x} \cdot \bs{x}_i(t) \right)$ as a sum of Green's functions of the Laplacian. 
The unit normal on the sphere is given by $\hat{\bs{k}} = \bs{x}/|\bs{x}|$. Because vorticity is materially conserved in the velocity field, the motion of point vortices is given by $\dot{\bs{x}}_i = \bs{u}(\bs{x}_i)$, i.e.,
\begin{equation*}
  \dot{\bs{x}}_i = \bs{x}_i \times \nabla \psi(\bs{x}_i) \quad i=1,2,\ldots,M,
  \label{eq:pvm}
\end{equation*}
where a unit sphere will assumed. The equations of motion may also be written as a Hamiltonian system with Lie-Poisson structure
\begin{equation}
  \Gamma_i \dot{\bs{x}}_i = \bs{x}_i \times \nabla_{\bs{x}_i} H \quad i=1,2,\ldots,M,
  \label{eq:pvs}
\end{equation}
where the Hamiltonian, defined by $H = \int_\mathcal{D} \frac{1}{2} |\bs{u}|^2 d\bs{x}$, is given by
\begin{equation*}
  H = -\sum_{i=1}^M \sum_{j=1}^{i-1} \frac{\Gamma_i \Gamma_j}{4 \pi} \ln \left( 2 - 2 \bs{x}_i \cdot \bs{x}_j \right).
  \label{eq:Hamiltonian}
\end{equation*}

By introducing $y = \left(\bs{x}_1^T, \bs{x}_2^T, \ldots, \bs{x}_M^T \right)^T$, equation \eqref{eq:pvs} can be written in the more compact form (\ref{HODE}) with the block-diagonal structure matrix
\begin{equation*}
  B(y) = 
  \begin{bmatrix}
    \Gamma_1^{-1} \widehat{\bs{x}}_1 &                              &        &   \\
                                 & \Gamma_2^{-1} \widehat{\bs{x}}_2 &        &   \\
                                 &                              & \ddots &   \\
                                 &                              &        & \Gamma_M^{-1} \widehat{\bs{x}}_M
  \end{bmatrix},
\end{equation*}
where $\widehat{\bs{x}}_i$ denotes the $3\times 3$ skew-matrix satisfying $\widehat{\bs{x}}_i \bs{a} := \bs{x}_i \times \bs{a}$, for all $\bs{a}\in\R^3$.
The Poisson bracket for the system is given equivalently by
\begin{align*}
  \poibra{F}{G} = \sum_{i=1}^M \frac{1}{\Gamma_i} \nabla_{\bs{x}_i} F \cdot (\bs{x}_i \times \nabla_{\bs{x}_i} G) \quad \text{or} \quad 
  \poibra{F}{G} = \nabla F(y)^T B(y) \nabla G(y).
\end{align*}
This Poisson structure is a generalization of the rigid body Poisson structure and also occurs in ferromagnetic spin lattices \cite{Faddeev:87,Frank04,FrHuLe97} and elastic rods (e.g.~\cite{KeMa97}).

The vortex positions are defined in Cartesian coordinates, but initial positions $\bs{x}_i(0)$ are chosen on the sphere. Because each $|\bs{x}_i|$ is a Casimir of the Poisson bracket it is ensured that the vortices remain on the sphere. This restricts the effective phase space of the system to the direct product of $M$ spheres $S^2$.  Furthermore, the rotational symmetry of the sphere gives rise to three Noether momenta, which are expressed by the angular momentum vector 
\begin{equation}
  \bs{J} = \sum_{i=1}^M \Gamma_i \bs{x}_i.
  \label{eq:Noether}
\end{equation}
When studying the statistics of point vortices on the disk, B\"{u}hler \cite{Buhler:06} did not observe the (planar) angular momentum to be of great importance. On the sphere, however, the angular momentum \emph{does} play an important role in the statistics.

The GBK thermostat (\ref{eq:GBK}) is only applicable to nondivergent systems $\nabla\cdot f\equiv 0$. It is straightforward to check that this condition holds for the spherical point vortex model.

\subsection{Time integration}
\label{sec:ti}
A numerical integrator can be constructed by splitting the differential equations into integrable subproblems (see related ideas in  \cite{Zhang:93,Patrick:00}).   We develop a new integrator for the system in  \ref{sec:2vortex} which exactly preserves all Casimir functions of the system.  Furthermore, backward error analysis for symplectic integrators can be extended to Poisson systems to explain approximate conservation of the Hamiltonian \cite{Hairer:06}.  Due to the additive form of the angular momentum vector (\ref{eq:Noether}), it may also be preserved exactly using a pairwise splitting.
By expanding the Hamiltonian into its pairwise terms in the dynamics we find
\begin{equation}
  \dot{y} = B(y) \nabla H(y)  =  \sum_{i<j} B(y) \nabla H_{ij}(\bs{x}_i,\bs{x}_j), \qquad 
  H_{ij} = \frac{\Gamma_i \Gamma_j}{8 \pi} \ln\left( 2 - 2\bs{x}_i \cdot \bs{x}_j \right)
  \label{eq:Hsplit}
\end{equation}
Each pairwise interaction is represented by the dynamical system $\dot{y} = B \nabla H_{ij}$ with the associated time-${\Delta t}$ flow map $\phi^{i,j}_{{\Delta t}}$. The time-${\Delta t}$ flow map of the dynamics $B \nabla H$ may be approximated by a symmetric composition of pair flows
\begin{equation}
  \Phi_{\Delta t} = \prod_{(i,j) \in C} \phi^{i,j}_{\Delta t/2} \circ \prod_{(i,j) \in C^*} \phi^{i,j}_{\Delta t/2},
  \label{eq:split}
\end{equation}
where $C$ is an ordered set of all possible pairs $(i,j)$ with $i<j$ and $C^*$ denotes the reverse ordering. This symmetric splitting yields a consistent numerical method of second order accuracy. The details of the integration procedure involving exact solution of the pairwise interaction is detailed in \ref{sec:2vortex}.

\begin{figure}
  \begin{center}
    \includegraphics[width=.49\textwidth]{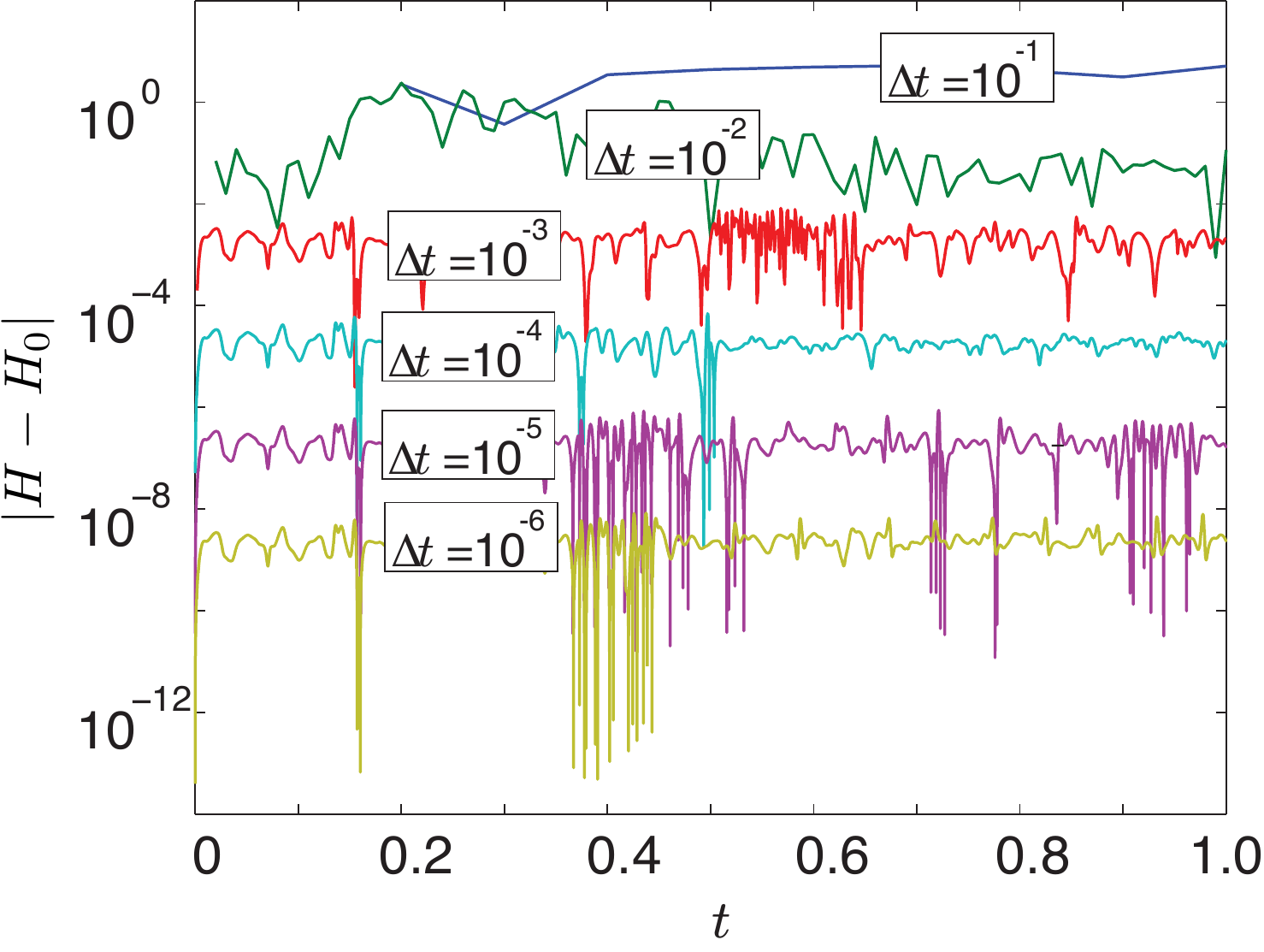}
    \includegraphics[width=.49\textwidth]{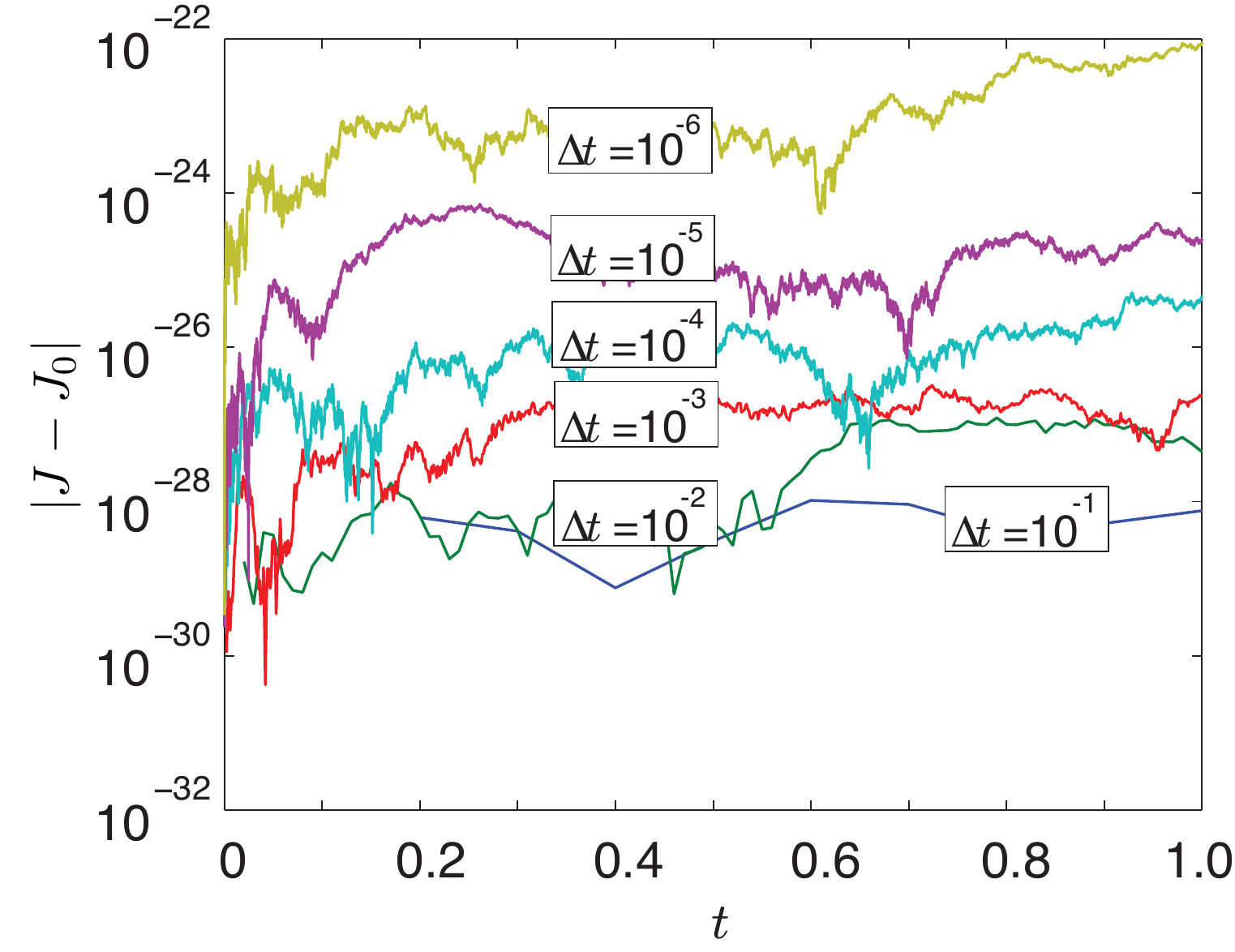}
  \end{center}
  \caption{Error in the energy (left) and angular momentum (right) for different time step sizes. Both from simulations with 16 vortices, of which 4 with strength $\pm 5$ and the rest with strength $\pm 1$. The momentum error is due to limited floating point accuracy. With decreasing time step the number of time steps increases and the inaccuracies accumulate, but they remain well within any reasonable demand for accuracy.}
  \label{fig:convH}
  \label{fig:convJ}
\end{figure}

Because the flow map of each vortex pair is the exact solution of the local Poisson system $\dot{y} = B(y) \nabla H_{ij}$ and also respects the Casimirs of the system, the composition $\Phi_{\Delta t}$ is a Poisson integrator \citep[p.~247]{Hairer:06}. Expanding the angular momentum vector as $\bs{J} = \bs{J}_{ij} + \sum_{k\neq i,j} \Gamma_k \bs{x}_k$ we note that the time integration of any pair $(i,j)$ preserves the local angular momentum $\bs{J}_{ij}$ and leaves the other vortices untouched. Hence the angular momentum is exactly conserved by the splitting method. The Hamiltonian is not exactly conserved under the motion of vortex pair, but
the error can be studied by backward error analysis; see e.g.~\cite{Hairer:06}. 
Figure \ref{fig:convH}a shows the error in the energy for simulations over a range of time step sizes, confirming second order convergence. The angular momentum should be conserved exactly by the Strang splitting. The results displayed in Figure \ref{fig:convJ}b confirm this as the errors are always well within machine precision.

\subsection{Thermostat perturbation vector}

There is flexibility in the choice of the perturbation vector field $g(y)$ in (\ref{eq:GBK}).  Its selection is determined both by the need for ergodicity with respect to the target measure and the need to preserve some invariants of the vector field $f(y)$.  We distinguish between invariants of $f$ whose values are known precisely, due to problem geometry for instance, and those whose values are uncertain and only known in expectation.  For point vortices on the sphere, the lengths of the vortex positions $|\bs{x}_i|$ are Casimir invariants, arising from the embedding of the sphere in $\R^3$, and are not subject to uncertainty. We choose a perturbation vector $g(y)$ that respects these structural invariants.

The double-bracket dissipation developed in \cite{Bloch:96} preserves Casimirs of the original system and is a candidate for $g(y)$:
\begin{equation}
  \tilde{g}_i(\bs{x}_i) = \sum_{j \neq i}\bs{x}_i \times \bs{x}_i \times \frac{\Gamma_j}{4\pi} \frac{\bs{x}_j}{1-\bs{x}_i \cdot \bs{x}_j}.
  \label{eq:double-bracket}
\end{equation}
The denominator in \eqref{eq:double-bracket} causes stiffness when like-signed vortices approach one another, restricting the step size of an explicit splitting method. To alleviate these matters we use a modified scheme defined by
\begin{equation}
  g_i(\bs{x}_i) = \sum_{j \neq i}\bs{x}_i \times \bs{x}_i \times \frac{\Gamma_j}{4\pi} \bs{x}_j.
  \label{eq:thermostat}
\end{equation}
The desirable properties of the thermostat are unaffected by this modification. \ref{sec:titd} contains a detailed description of the numerical integration of these dynamics.

The thermostat (\ref{eq:GBK}) is designed to sample a target density $\rho(y) \propto e^{-A(y)}$ on the phase space of $y$. The thermostat variable $\xi$ is normally distributed, yielding the extended distribution $\rho \propto e^{-A(y) - \frac{1}{2}\xi^2}$. 
The perturbation vector field $g$ must additionally ensure that the thermostated system is ergodic in the target density. 
Because the target measure is positive for all open sets on the phase space, hypoellipticity of the Fokker-Planck equation associated with \eqref{eq:GBK} is sufficient to prove uniqueness of the invariant measure \citep{BaFrLe:13}. Hypoellipticity follows from H\"ormander's controllability condition \citep{Rey-Bellet:06}. The condition has been tailored to GBK thermostats in \cite{BaFrLe:13}, but it is difficult to check in practice. Here we instead check empirically that single trajectories have statistics that agree with the target distribution.

\subsection{Maximum entropy model}
\label{sec:sm}

To apply the methodology proposed in Section \ref{sec:Method} in the setting of a reduced model for point vortices, we  use point vortices distributed evenly over the surface of the sphere as the prior $\pi$.
it remains to specify for which functions of the conserved quantities $H$ and $\bs{J}$ the expectations will be observed during simulation of the full model. 

In \cite{Dubinkina:10} a thermostat was used to model a set of point vortices on a disk in the canonical ensemble. To accurately reproduce statistics from a full model with a moderate number of point vortices, it was necessary to modify the canonical density with a term quadratic in the Hamiltonian, that is, a density of the form $\rho(y) \propto \exp(-\beta H(y) - \gamma H(y)^2)$.  Motivated by the experience in \cite{Dubinkina:10}, we choose observations that include linear and quadratic functions of $H$ and $\bs{J}$.

If the angular momentum of the full system is zero, then there is no directional preference for the angular momentum vector $\bs{J}$. We consider the following set of observables:
\begin{equation}\label{eq:obs}
	C_1 = H, \quad 
	C_2 = |\bs{J}|^2, \quad
	C_3 = H^2, \quad
	C_4 = |\bs{J}|^4, \quad
	C_5 = H^2|\bs{J}|^2,
\end{equation}
and denote the corresponding Lagrange multipliers by $\beta_H$, $\beta_J$, $\gamma_H$, $\gamma_J$, $\gamma_{HJ}$.

The least-biased density consistent with observations of the $\mathbb{E} C_k$ is	
\begin{equation}
  \tilde{\rho}(H)=e^{-\beta_H H -\beta_J |\bs{J}|^2 -\gamma_H H^2 -\gamma_J |\bs{J}|^4 - \gamma_{HJ} H |\bs{J}|^2}.
  \label{eq:canonical_finite}  
\end{equation}

\section{Numerical comparison\label{sec:numcom}}
To verify the methodology proposed in this article for correcting expectations, we apply it to a reduced model of point vortices similar to the configuration used in \cite{Buhler:02,Dubinkina:10}.  We distinguish between three models.  The \emph{full model} consists of a system (\ref{eq:pvs}) of $M_\text{full}=288$ point vortices, of which 8 strong vortices of circulation $\Gamma_j =\pm1$ and 280 weak vortices of circulation $\Gamma_j=\pm\frac{1}{5}$. Both strong and weak classes are comprised of equal numbers of positively and negatively oriented point vortices.
The \emph{reduced model} consists of (\ref{eq:pvs}) with just $M=8$ strong vortices.  Finally, the \emph{corrected model} consists of a thermostated system (\ref{eq:GBK}) with unperturbed vector field $f$ given by (\ref{eq:pvs}) for $M=8$ strong vortices, perturbation vector field $g$ given by (\ref{eq:thermostat}),  and equilibrium measure defined by the least-biased density (\ref{eq:canonical_finite}).  Additionally, we compare with Metropolis-Hastings samples from the least-biased density (\ref{eq:canonical_finite}) to help distinguish between errors incurred due to the maximum-entropy model and those due to the thermostat.

We run seven long simulations of the full model with angular momentum vector $\bs{J}_\text{full}=\bs{0}$ and total energies chosen from the set $H_\text{full} \in \{ -2, -1, 0, 1, 2 \}$.  
For each run we determine the time averages of the observables (\ref{eq:obs}) for the subset of strong vortices. When computing the Hamiltonian $H$ we include only the internal coupling between strong vortices.  The time averages are tabulated in Table \ref{tab:obs_compar}).

\begin{table}[hpt]
\caption{Full model observations and (in parentheses) corrected values of first integrals \label{tab:obs_compar}}
 \begin{center}
\begin{tabular}{l|r|r|r|r|r}
& $\langle H \rangle$ & $\langle |\bs{J}|^2 \rangle$ & $\langle H^2 \rangle$ & $\langle |\bs{J}|^4 \rangle$ & $\langle H|\bs{J}|^2 \rangle$ \\\hline
\textbf{$H_\text{full} = -2$}&-0.33 (-0.38)&4.59 (4.45)&0.22 (0.23)&-0.63 (-0.98)&34.58 (31.45)\\
\textbf{$H_\text{full} = -1$}&-0.11 (-0.18)&4.78 (4.68)&0.10 (0.12)&0.38 (-0.01)&37.55 (35.88)\\
\textbf{$H_\text{full} = 0$}&0.02 (-0.04)&4.63 (4.56)&0.08 (0.08)&0.90 (0.60)&35.26 (34.30)\\
\textbf{$H_\text{full} = 1$}&0.17 (0.15)&4.74 (4.75)&0.13 (0.12)&1.73 (1.61)&37.76 (37.44)\\
\textbf{$H_\text{full} = 2$}&0.31 (0.28)&4.87 (5.00)&0.22 (0.21)&2.49 (2.46)&39.26 (41.74)\\
\end{tabular}
\end{center}
\end{table}

Given the time averages, we compute the Lagrange multipliers using the algorithm described in Section \ref{sec:adaptive} with prior distribution $\pi$ the uniform distribution on the sphere.
The Lagrange multipliers are also recorded in Table \ref{tab:tabHs}. The magnitude of $\gamma_{\lbrace H,J,HJ\rbrace}$ indicates that all observations are relevant for all but the most negative energy levels.

Subsequently, we run simulations of the corrected model using the computed parameters. 
Table \ref{tab:obs_compar} also records expectations from the thermostat-corrected model.

\begin{table}[hpt]
  \caption{Lagrange multipliers for each energy level.}
  \begin{center}
    \input{tabHs_Lagrange}
  \end{center}
  \label{tab:tabHs}
\end{table}

By analogy with canonical statistical mechanics, we may think of the weak vortices that are ignored in the reduced model as forming a reservoir with which our reduced model exchanges energy and angular momentum.
Experience with canonical statistical mechanics of point vortices in the plane \cite{Buhler:02,Dubinkina:10} suggests that for small reservoir sizes the canonical Gibbs distribution must be modified with higher order terms to agree with the full system statistics.  Table \ref{tab:tabNs} contains a study of the Lagrange multipliers as a function of the full system size $M_\text{full}$, confirming that the Lagrange multipliers $\gamma_H$, $\gamma_J$ and $\gamma_{HJ}$ are more significant for smaller $M_\text{full}$.  

\begin{table}[hpt]
  \caption{Lagrange multipliers as a function of $M_\text{full}$, all for $H_\text{full} = 0$.}
  \begin{center}
    \input{tabNs_Lagrange} 
  \end{center}
  \label{tab:tabNs}
\end{table}

The energy of the strong vortices may become arbitrarily large because of the singularity in the Hamiltonian as two vortices approach each other. But the same is true for the energy in the reservoir. If there are at least three reservoir vortices and not all those vortices have the same sign, the reservoir can supply or remove any amount of energy.

The condition on the angular momentum is more interesting. The system of strong vortices, all with strength $\pm \Gamma_\mathrm{strong}$, has angular momentum satisfying $|\bs{J}_\text{red.}| \leq M\Gamma_\mathrm{strong}$. For the reservoir it holds that $|\bs{J}_\mathrm{weak}| \leq (N-M)\Gamma_\mathrm{weak}$. It is necessary that the reservoir can supply sufficient angular momentum, that is
\begin{align*}
  M\Gamma_\mathrm{strong} \leq (M_\text{full}-M)\Gamma_\mathrm{weak} \Leftrightarrow \frac{M_\text{full}-M}{M} \geq \frac{\Gamma_\mathrm{strong}}{\Gamma_\mathrm{weak}}.
  \label{eq:neccond}
\end{align*}
In the thermal bath simulations discussed in this section $M=8$ and $\Gamma_\text{strong}/\Gamma_\text{weak}=5$, this means $M_\text{full}$ should satisfy $M_\text{full} \geq 48$. The smallest system considered ($M_\text{full}=36$) does not, explaining its eccentric parameter values in Table \ref{tab:tabNs}.

\subsection{Equilibrium results}
In this section we compare statistical properties of the corrected model with those of the full and reduced models.
In Figures \ref{fig:res_lowest}--\ref{fig:res_highest} we show histograms of a number of solution features for the 8 vortex model: the distributions of $H$ and $|\bs{J}|$, as well as typical distances between like- and opposite-signed vortices, a metric used by B\"uhler \cite{Buhler:02}. In each histogram, the statistics corresponding to the strong vortices in the full model, the reduced model, thermostat-corrected reduced model, and Metropolis-Hastings samples are displayed.  Figures \ref{fig:res_lowest}--\ref{fig:res_highest} correspond to approximate total energies $H_\text{full} \approx -2$, 0 and 2, respectively.

The full and reduced model simulations are performed with a time step of $5 \times 10^{-3}$ and run up to $T=5 \times 10^4$, taking $10^5$ samples spaced evenly in time. For the Metropolis-Hastings method we use $10^6$ samples. 
The same figures also show results from the thermostated system (dash-dot lines), run with a time step of $10^{-3}$ up to $T=10^6$, taking $10^6$ samples.  The parameters in \eqref{eq:GBK} were set to be $\epsilon = 10$ and $\gamma = 0.1$. These results confirm that the thermostated system samples the least-biased density closely.

The reduced model is Hamiltonian and the Poisson integrator ensures that the energy is conserved with a standard deviation of order $10^{-3}$ and the angular momentum constant to machine precision. Both cases correspond to approximate delta-distributions in the upper histograms in Figures \ref{fig:res_lowest}--\ref{fig:res_highest}.  Note that due to the high skewness of the distribution for $|\bs{J}|$, the observed mean differs significantly from the median and mode, implying some ambiguity in choosing the angular momentum for  an appropriate initial condition for the reduced model.

\newcommand{\myfiguretext}{
The upper left and right panels compare strong vortex energy and angular momentum magnitude. The lower left (resp.~right) panel compares the distance between like (resp.~opposite) signed strong vortices. The parameters are specified in the text.
}
\newcommand{\myfiguretexttwo}{
Same panel layout as Figure \ref{fig:res_lowest}.
}
\begin{figure}[hb!]
	\begin{center}
		\includegraphics[width=\figwidth\textwidth]{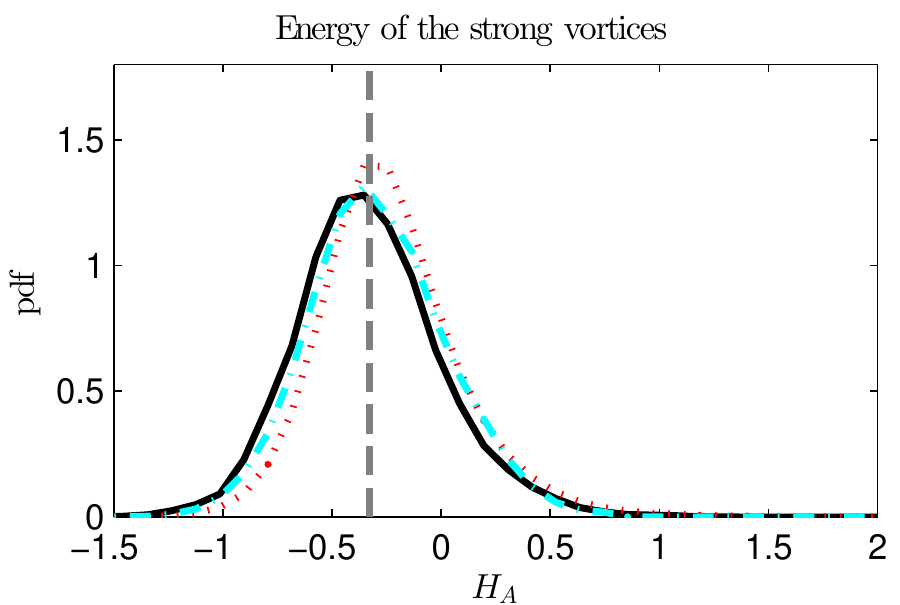} 
		\includegraphics[width=\figwidth\textwidth]{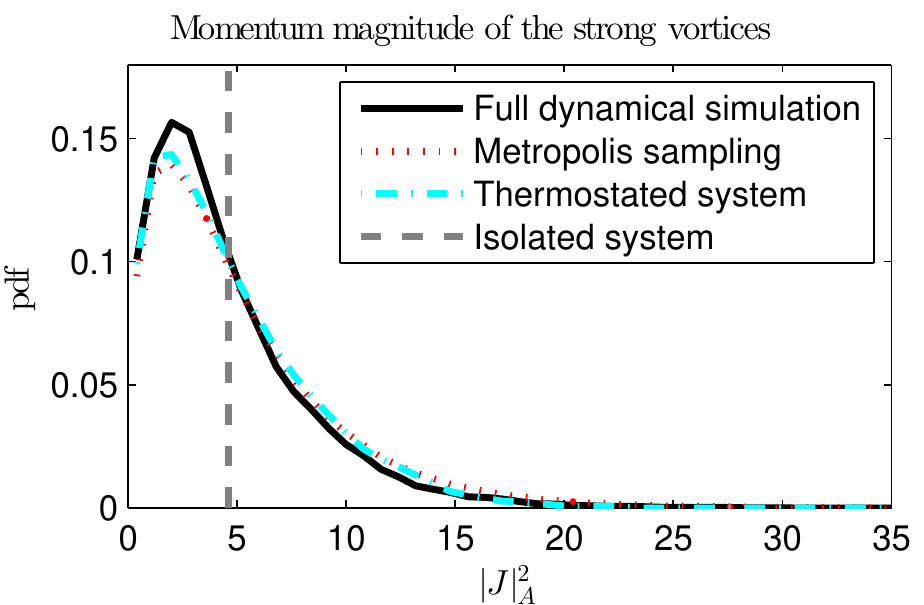}\\
		\includegraphics[width=\figwidth\textwidth]{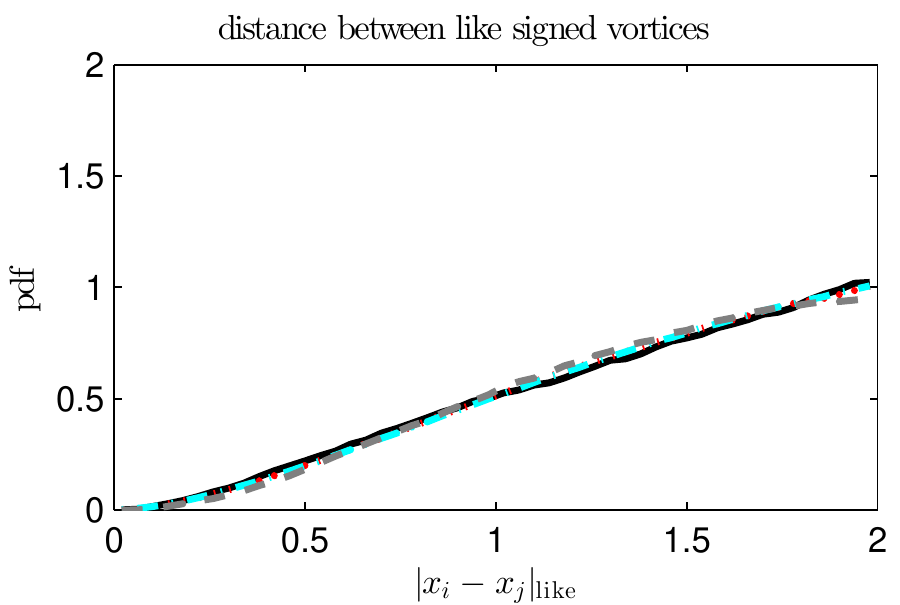} 
		\includegraphics[width=\figwidth\textwidth]{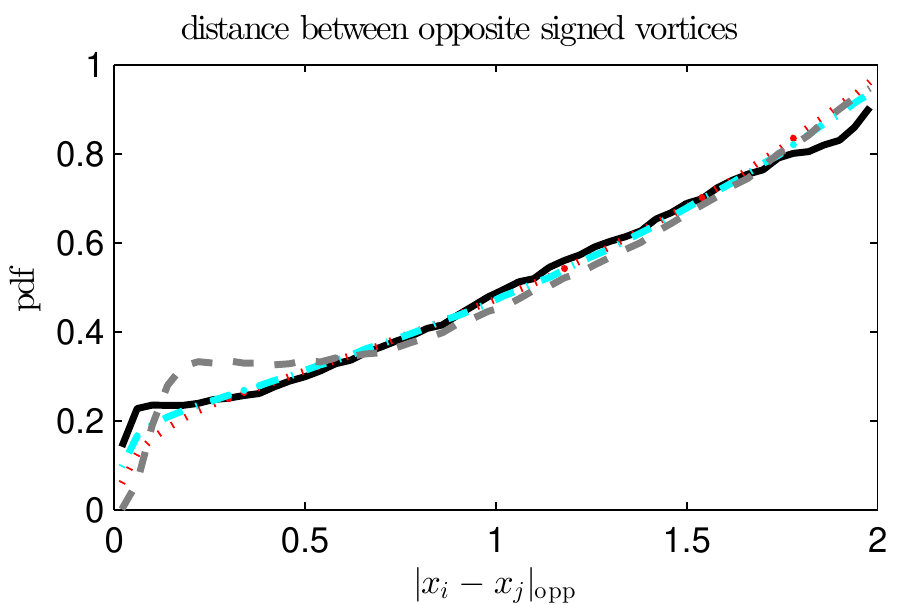} 
	\end{center}%
	\caption{Histograms for $H_\text{full} \approx -2$. \myfiguretext}
	\label{fig:res_lowest}
\end{figure}
%

%
\begin{figure}[hp!]
	\begin{center}
		\includegraphics[width=\figwidth\textwidth]{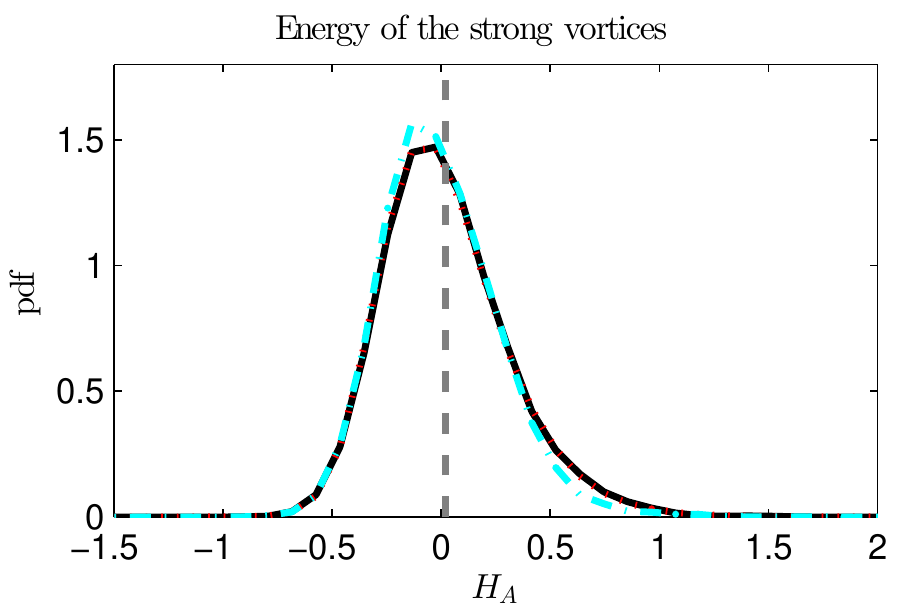} 
		\includegraphics[width=\figwidth\textwidth]{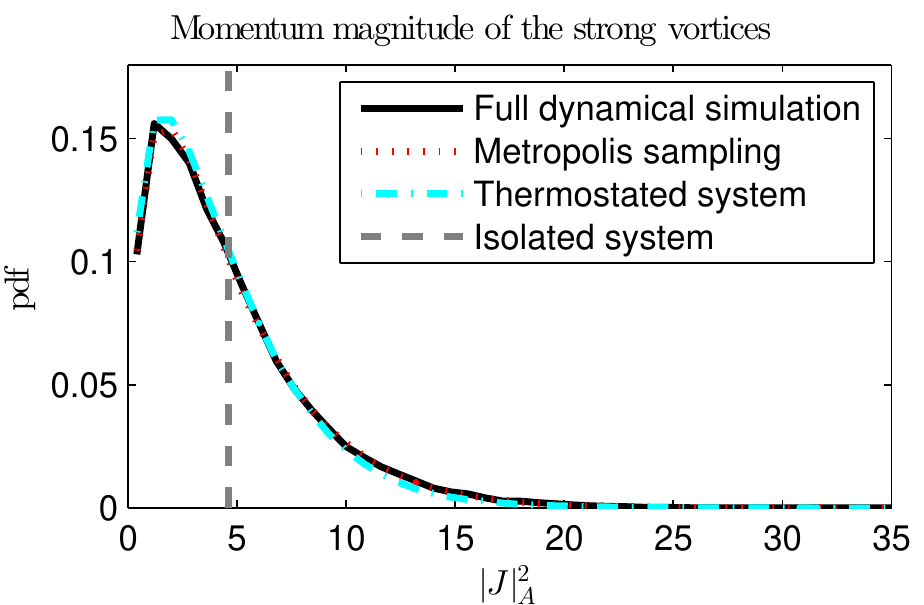}\\
		\includegraphics[width=\figwidth\textwidth]{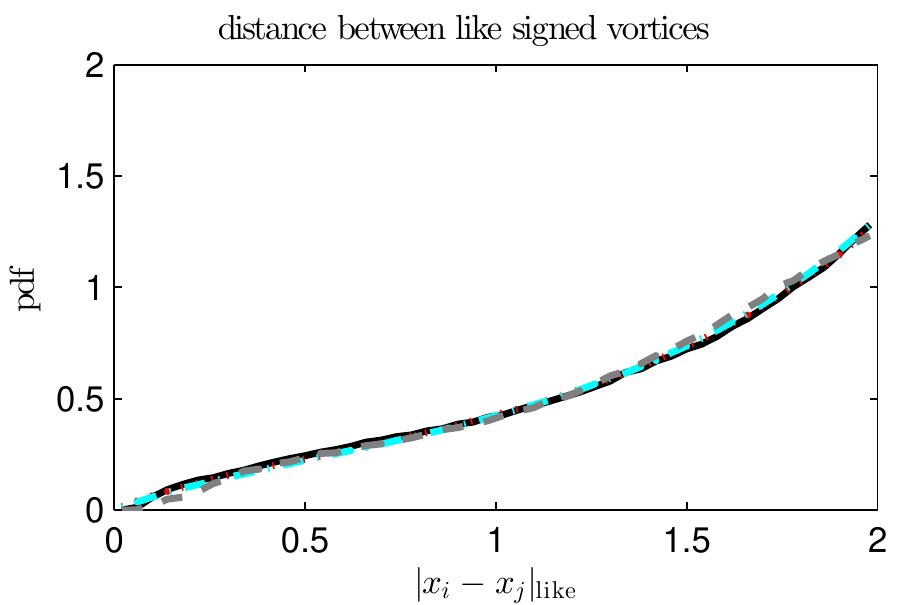}
		\includegraphics[width=\figwidth\textwidth]{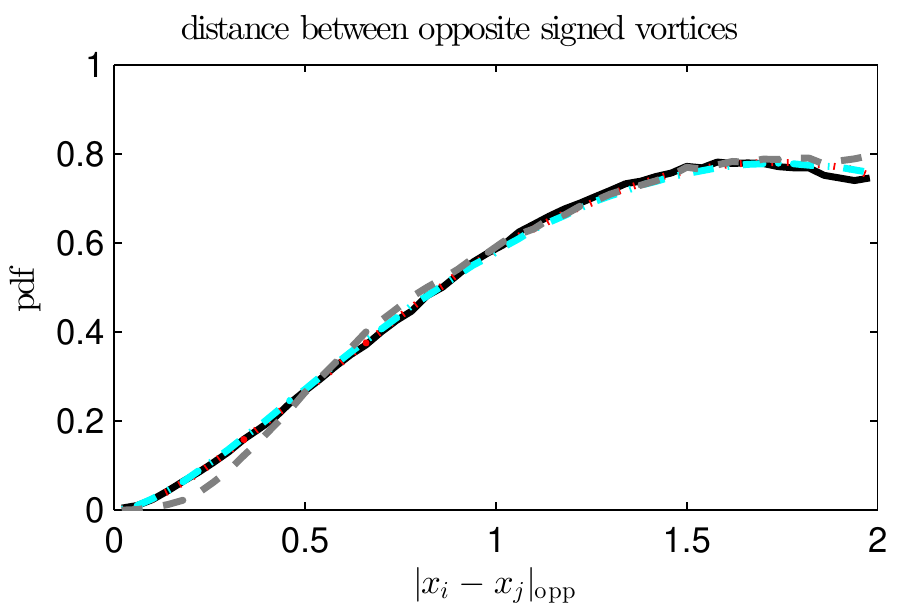} 
	\end{center}%
	\caption{Histograms for $H_\text{full} \approx 0$. \myfiguretexttwo}
	\label{fig:res_neutral}
\end{figure}
%
%
%
\begin{figure}[hp!]
	\begin{center}
		\includegraphics[width=\figwidth\textwidth]{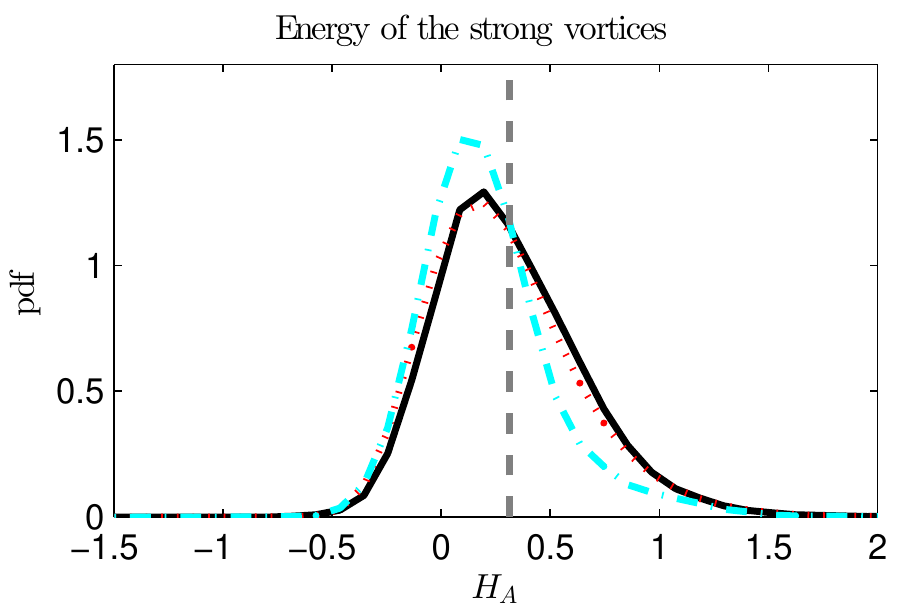} 
		\includegraphics[width=\figwidth\textwidth]{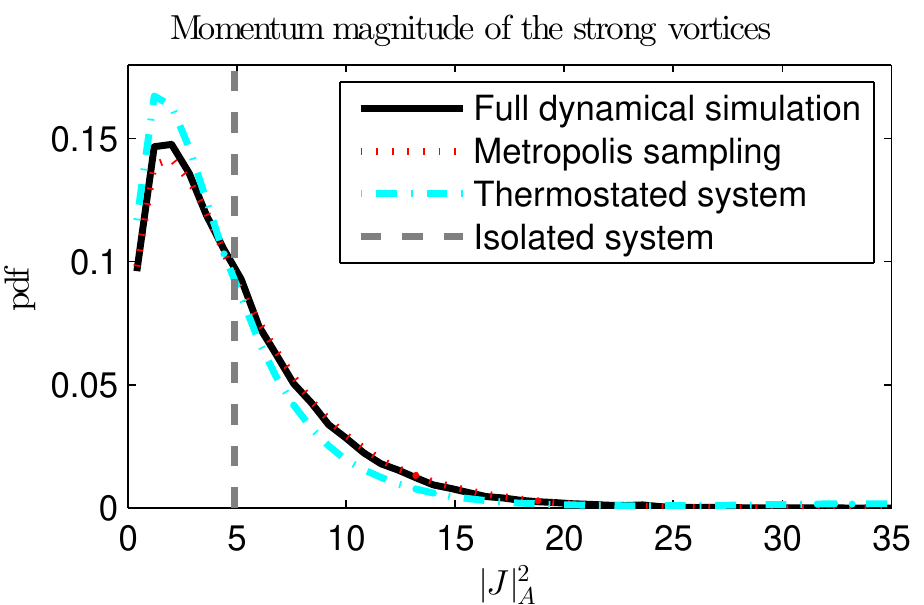} \\
		\includegraphics[width=\figwidth\textwidth]{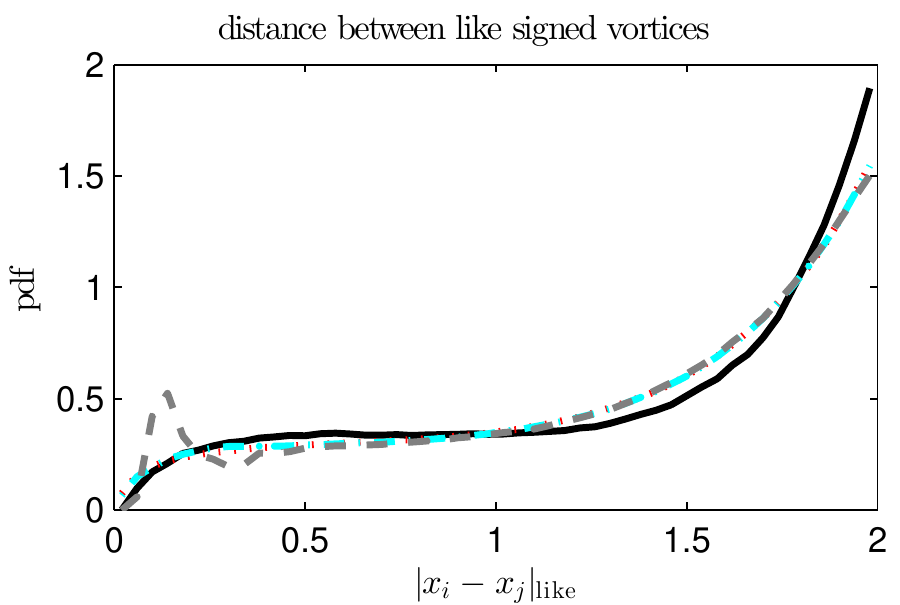}
		\includegraphics[width=\figwidth\textwidth]{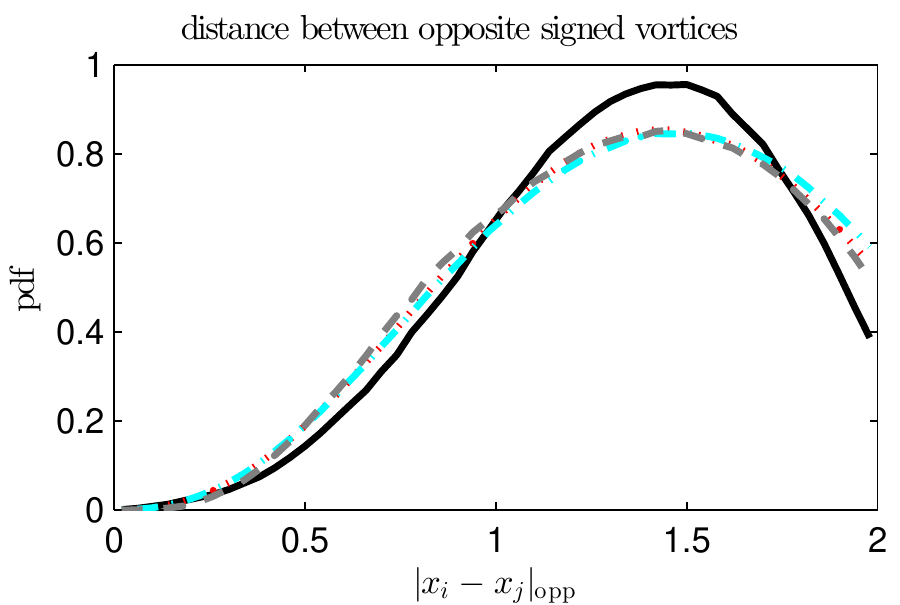} 
	\end{center}%
	\caption{Histograms for $H_\text{full} \approx 2$. \myfiguretexttwo}
	\label{fig:res_highest}
\end{figure}

A simple Hamiltonian reduced model is naturally incapable of sampling the energy and angular momentum spectra, since these quantities are first integrals. In turn, the reduced model shows significant bias in statistics such as vortex separation.  The thermostat-corrected model faithfully samples the least-biased probability density, as indicated by the good agreement in the histograms of the corrected model and Metropolis-Hastings samples.  The least-biased density does a good job of approximating the strong-vortex statistics in the negative to moderate total energy regime.  At large positive total energies, the strong vortex energy and angular momentum distributions are still well-represented by the least-biased PDF, but some bias in the vortex separations can be observed. The closeness of the thermostat results to those from the Metropolis-Hastings sampling indicate the error lies in the choice of least-biased density, not in the thermostat sampling.

\subsection{Dynamic consistency}
The results in the previous section confirm that the thermostated simulations lead to equilibrium distributions of observables $H$ and $|\bs{J}|$ similar to those of the full system. In this section we address the issue of the degree to which our equilibrium correction mechanism disturbs dynamics, as encoded in autocorrelation functions and diffusivity.  Diffusivity was considered by \cite{Chavanis:01} for a system of identical point vortices and by \cite{Cotter:09} for a wide array of problems with scale separation. We emphasize that the values of the thermostat parameters $\varepsilon$ and $\gamma$ have no impact on the equilibrium statistics presented in the previous section, and only affect the rate at which the least-biased PDF is sampled. Faster convergence to the equilibrium distribution correlates with a larger deviation from the unperturbed dynamics and vice-versa.

\subsubsection{Autocorrelation functions}

Given a sequence of $L$ equally spaced observation times $t_i \in [0,T]$ for $i \in [0,L]$, and the values of the relevant observable (in our case vortex position) $u_i=u(t_i)$ at those times, the discrete autocorrelation function is defined by
\begin{equation*}
  \label{eq:af}
  \nu^u_i = \frac{1}{L-i} \sum_{j=i}^L u(t_j) u(t_{j-i}).
\end{equation*}
A normalized autocorrelation function $\hat{\nu}^u$ is given by dividing each $\nu^u_i$ by $\nu^u_0$, i.e.~$\hat{\nu}^u_i = \nu^u_i/\nu^u_0$.

We average the autocorrelation function over all $3M$ (strong) vortex coordinates. 
Three symmetries in the problem justify this averaging: the vortex numbering is arbitrary; the choice of reference frame is arbitrary and the sign of the vortices appears in the dynamics as a reversal of time, to which the autocorrelation is insensitive.  Additionally, the observables $H$ and $|\bs{J}|$ are isotropic.

Furthermore we ensure that the phase space is well sampled by averaging the autocorrelation functions over an ensemble of $P$ solutions. The choice of ensemble initial condition is detailed in Section \ref{af:res} below. We then find the average autocorrelation function
\begin{align}
  \label{eq:avgaf}
 \nu_i 
&= \frac{1}{3MP(L-i)} \sum_{p=1}^P \sum_{m=1}^M \sum_{j=i}^L x^p_m(t_j) x^p_m(t_{j-i}) + y^p_m(t_j) y^p_m(t_{j-i}) + z^p_m(t_j) z^p_m(t_{j-i})
\end{align}
and the normalized average autocorrelation function
\begin{align}
  \label{eq:avgafnorm}
  \hat{\nu}_i 
&= \frac{1}{MP(L-i)} \sum_{p=1}^P \sum_{m=1}^M \sum_{j=i}^L \bs{x}^p_m(t_j) \cdot \bs{x}^p_m(t_{j-i}),
\end{align}
where a superscript $p$ represents the solution from ensemble member $p$.
The normalized autocorrelation function in \eqref{eq:avgafnorm} follows from the Casimirs $C_i = \bs{x}_i(t) \cdot \bs{x}_i(t) = 1 \,\forall\, i,t$.

\subsubsection{Diffusivity}
For general multiscale dynamical systems with a separation of slow and fast dynamics, it is often desirable to model fast forces by a diffusion process, resulting in stochastic differential equation of the form  \citep{Pavliotis:08,Hornung:97}
\begin{equation*}
  d X = f(X) d t + K(X) d W,
\end{equation*}
where $f$ represents the slow dynamics, $W$ is a Wiener process and $K$ is the diffusivity. The value of the diffusivity $K$ can be estimated by sampling solutions to the original, multiscale, problem and applying Kubo's formula
\begin{equation*}
  K_{\Delta \tau} = \frac{\ensbra{\Delta X \Delta X}}{2 \Delta\tau},
\end{equation*}
where $\Delta X$ represents displacement during the sampling interval $\Delta\tau$. Choosing the correct sampling interval is a notorious problem;  for a comparison see \cite{Pavliotis:08}. 

If we take the average diffusivity for each vortex coordinate we find
\begin{align*}
  K_{\Delta\tau} 
&= \frac{1}{6M\Delta\tau} \sum_{m=1}^M \ensbra{\Delta x_m \Delta x_m + \Delta y_m \Delta y_m + \Delta z_m \Delta z_m} \\
&= \frac{1}{6M\Delta\tau} \sum_{m=1}^M \ensbra{\Delta \bs{x}_m \cdot \Delta \bs{x}_m}.
\end{align*}
We assume the observations are given at the same times $t_i$ as before and that the sampling time is an integer multiple of the observation interval, i.e. $\Delta\tau = i \frac{T}{L}$. With an ensemble of $P$ simulations the diffusivity estimator would then be
\begin{align*}
  K_{\Delta\tau} 
&= \frac{1}{6MP\Delta\tau} \sum_{p=1}^P \sum_{m=1}^M \left( \bs{x}^p_m(t_i) - \bs{x}^p_m(0) \right) \cdot \left( \bs{x}^p_m(t_i) - \bs{x}^p_m(0) \right) \\
&= \frac{1}{6MP\Delta\tau} \sum_{p=1}^P \sum_{m=1}^M 2 - 2 \bs{x}^p_m(0) \cdot \bs{x}^p_m(t_i) \\
&= \frac{1}{3 \Delta\tau} - \frac{1}{3MP\Delta\tau} \sum_{p=1}^P \sum_{m=1}^M \bs{x}^p_m(0) \cdot \bs{x}^p_m(t_i),
\end{align*}
where again a superscript $p$ denotes the solution from ensemble member $p$. 
Averaging over all time series data yields the estimator:
\begin{align*}
  K^i_{\Delta\tau} 
&= \frac{1}{3\Delta\tau} - \frac{1}{3MP(L-i)\Delta\tau} \sum_{p=1}^P \sum_{j=i}^L \sum_{m=1}^M \bs{x}^p_m(t_{j-i}) \cdot \bs{x}^p_m(t_j) \\
&= \frac{1 - \hat{\nu}_i}{3 \Delta\tau}.
\end{align*}
This \emph{shift-averaged estimator} is shown in \cite{Cotter:09} to improve the quality of the estimator.

\subsubsection{Dynamical results}
\label{af:res}
\begin{figure}[htb!]
	\begin{center}
		\includegraphics[width=0.6\textwidth]{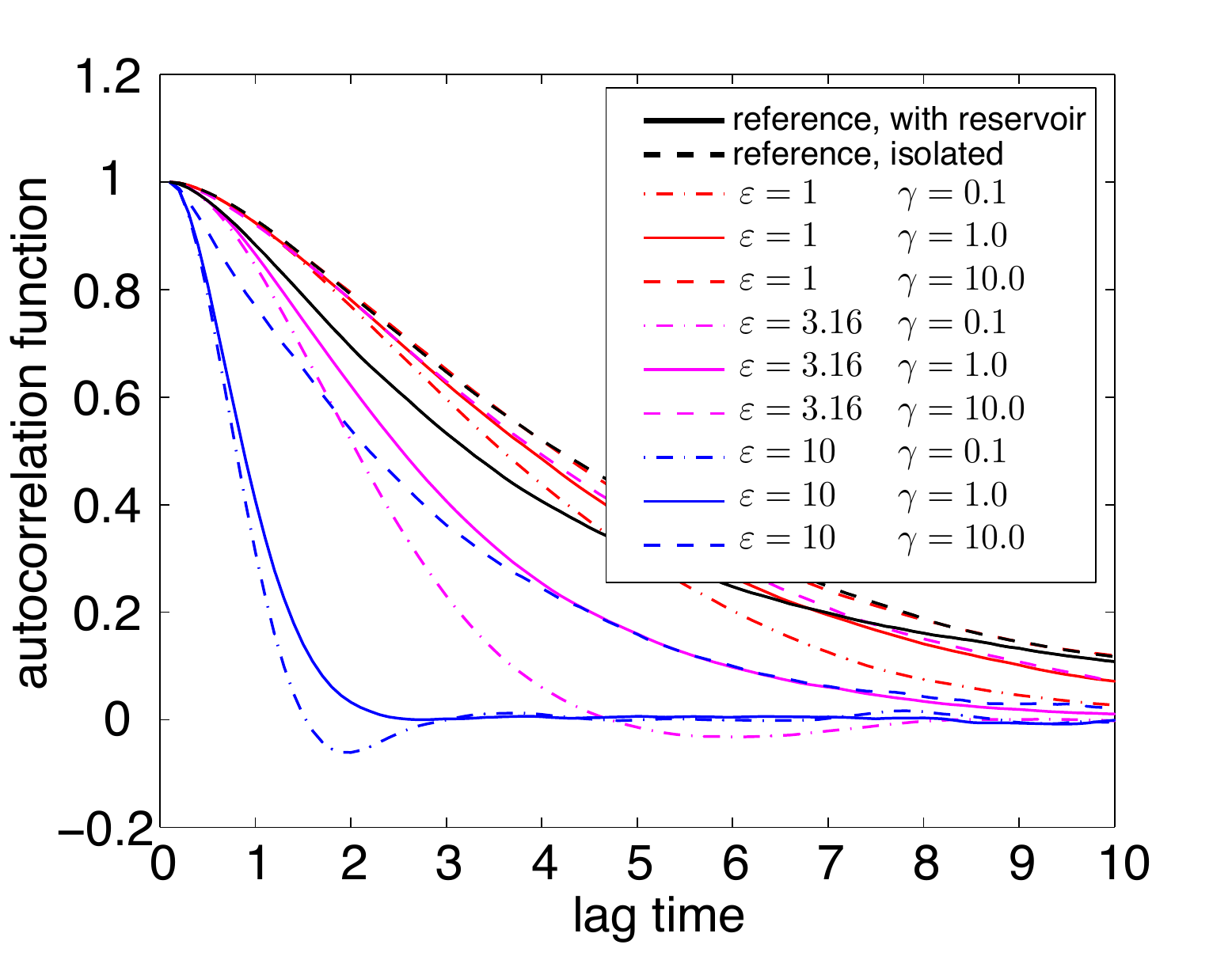} 
	\end{center}
	\caption{Comparison of autocorrelation of the vortex coordinates. The bold lines are two reference cases: the full model (solid) and the reduced model (dashed). The thin lines indicate autocorrelation functions of the thermostat-corrected model for indicated values of parameters $\varepsilon$ and $\gamma$.}
	\label{fig:cx}
\end{figure}

In Figure  \ref{fig:cx} we compare auto-correlation functions for the strong vortices in the full and reduced models as well as for the thermostat-corrected model over a range of parameters $\varepsilon$ and $\gamma$.
The thick solid black line represents the result for an (unthermostated) system in contact with $280$ weak ($\Gamma_B = \pm \frac{1}{5}$) vortices, with a total energy $H_\text{full}=0$. The results present the average over an ensemble of 1000 runs. For each simulation the initial placement of each strong vortex was taken uniformly over the sphere and the weak vortices were placed such that the full system satisfied $H_\text{full}=0$ and $\bs{J}_\text{full}=0$. The thick dashed black line represents the results for an ensemble of simulations of the isolated system, with everything else unchanged. The other lines represent results for thermostated simulations using the parameters as given in Table \ref{tab:tabHs} for the case of $H=0$.

\begin{figure}[htb!]
         \centering
	 \begin{minipage}[b]{.5\linewidth}
                 \centering
                 \includegraphics[width=\textwidth]{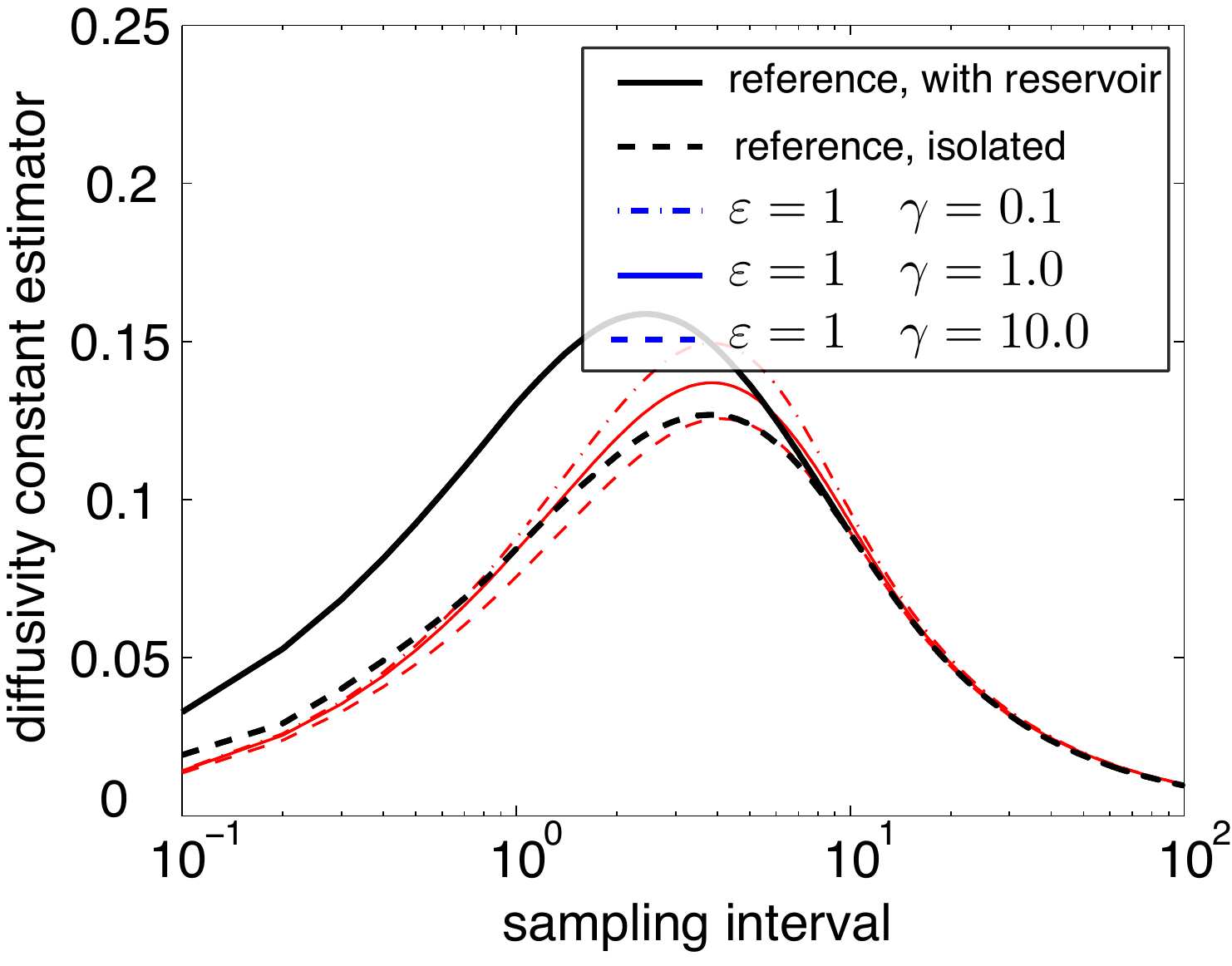}
                 \subcaption{}\label{fig:diff_e0}
         \end{minipage}%
         \begin{minipage}[b]{0.5\textwidth}
                 \centering
                 \includegraphics[width=\textwidth]{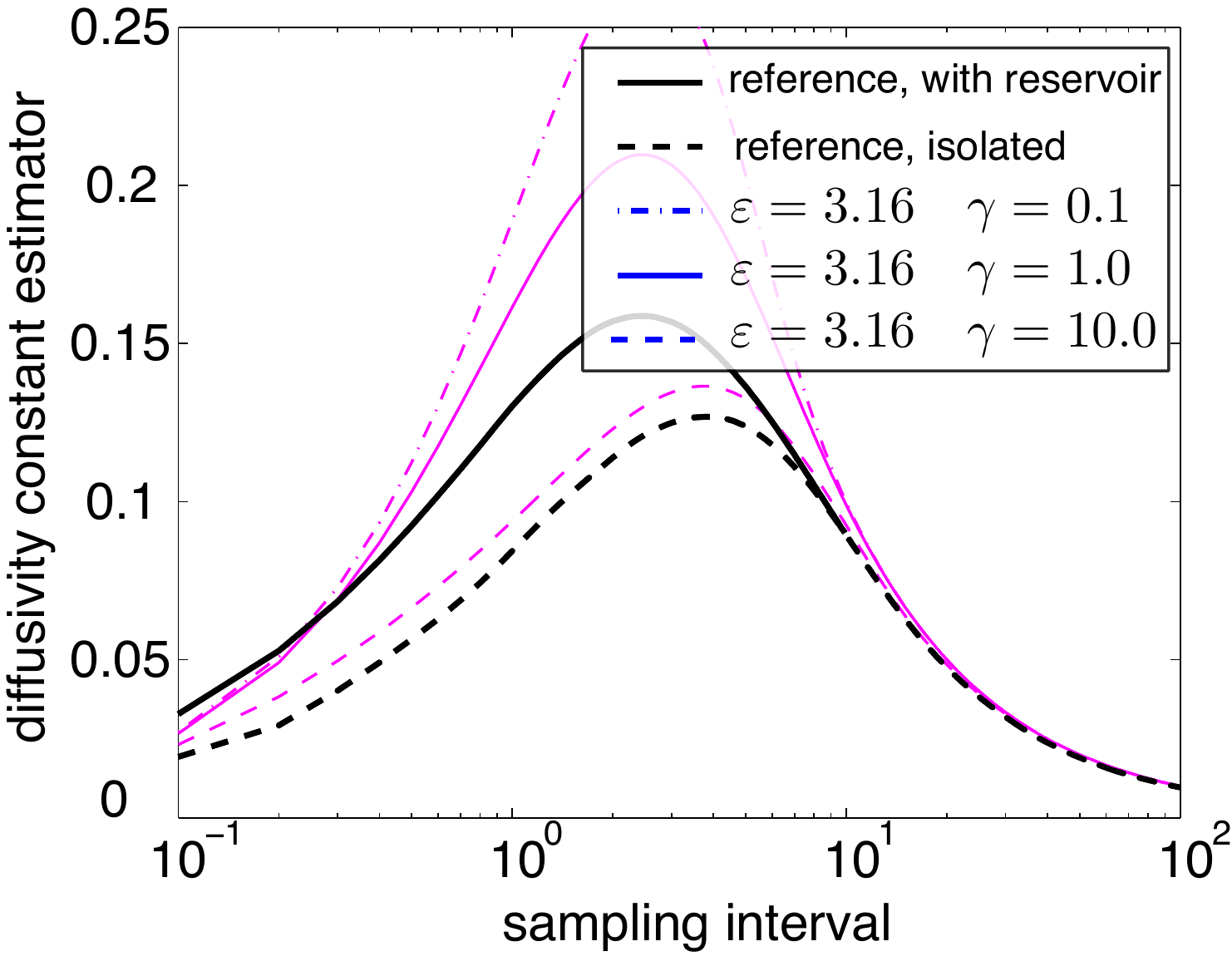}
                  \subcaption{}
                 \label{fig:diff_e05}
         \end{minipage}%
 
         \begin{minipage}[b]{0.5\textwidth}
                 \centering
                 \includegraphics[width=\textwidth]{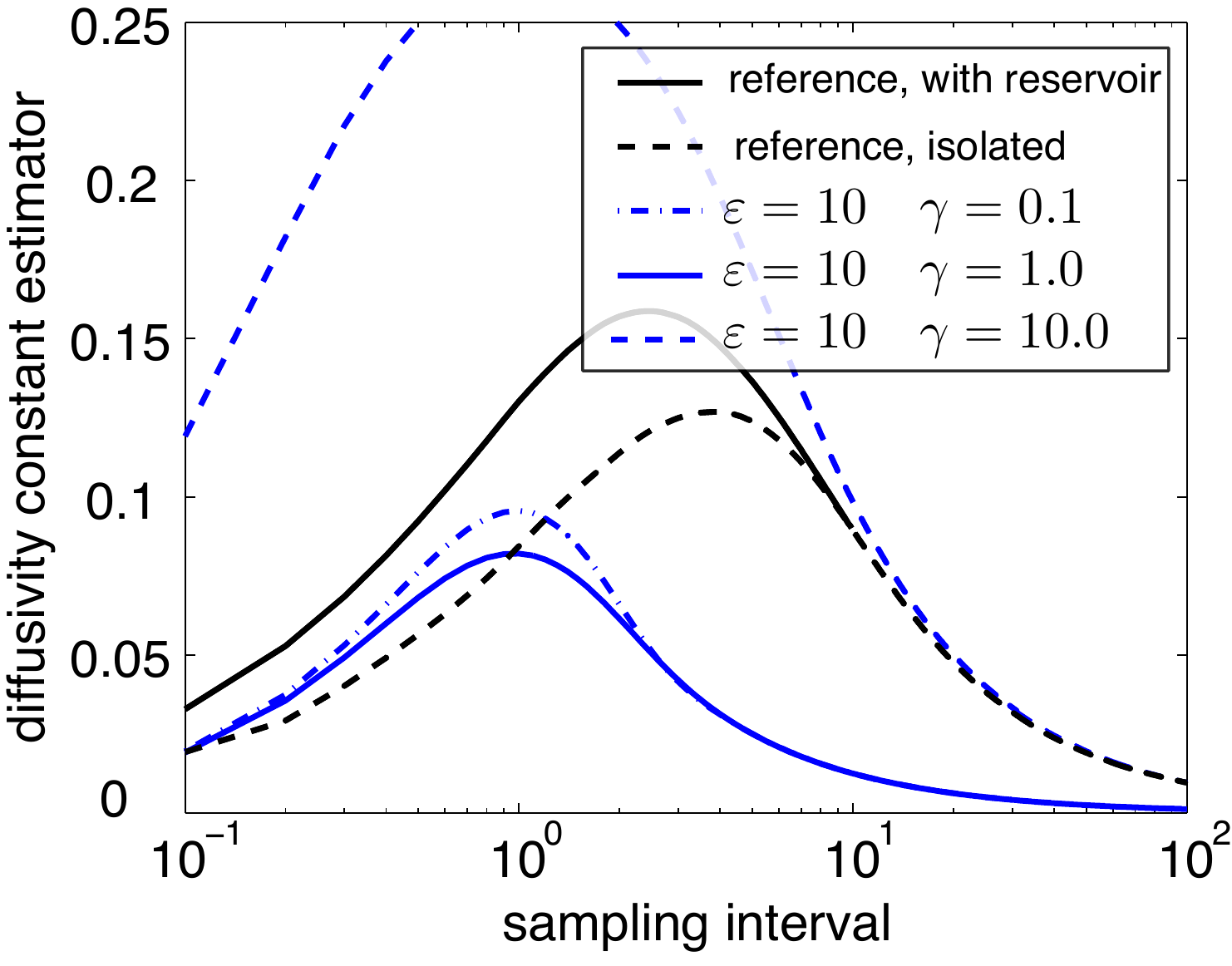}
                  \subcaption{}
                 \label{fig:diff_e1}
         \end{minipage}%
         \begin{minipage}[b]{0.5\textwidth}
                 \centering
                 \includegraphics[width=\textwidth]{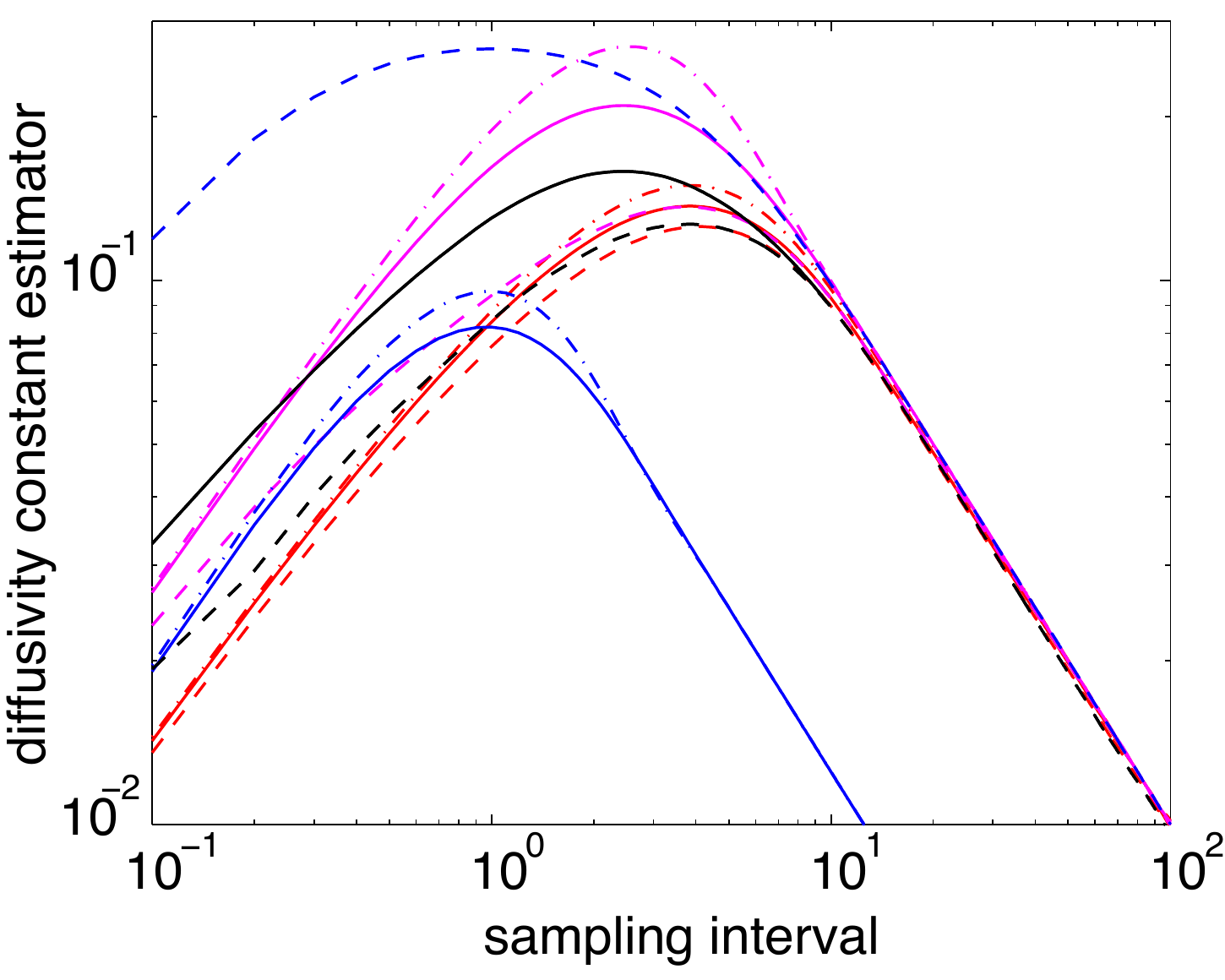}
                 \subcaption{}
                 \label{fig:diffloglog}
         \end{minipage}
  	\caption{Comparison of average diffusivity constant as a function of sampling intervals. In all figures, the bold lines indicate two reference cases: the full model (solid) and the reduced model (dashed). Subfigures (a), (b) and (c) show thermostated simulation results for $\epsilon$ equal to $10^0$, $10^{0.5}$ and $10^1$ respectively. The value of $\gamma$ is represented by dash-dot ($10^{-1}$), solid ($10^0$) or dashed ($10^1$) lines. A combined log-log plot of all parameter values is given in subfigure (d).}
 	\label{fig:diff}
\end{figure}

The corresponding diffusivity constants are presented in Figure \ref{fig:diff}. The results are taken from the same simulations as described in the paragraph above. Because this figure is visually more striking, we shall limit our discussion to the diffusion constant. 

For $\varepsilon$ small, the thermostat perturbation is weak, and both autocorrelation functions and diffusivity approach those of the reduced model with constant $H$, $\bs{J}$.  Also, the autocorrelations are insensitive to the parameter $\gamma$ in this regime.  For even smaller $\varepsilon$ the autocorrelations and diffusivities become indistinguishable from those of the reduced model.  Hence even though the dynamics samples the least-biased density on long time scales, its short time dynamics is similar to an unperturbed model.  For moderate $\varepsilon$, dependence on $\gamma$ becomes more pronounced, and a diffusivity closer to that of the full model can be achieved.  For even larger values of $\varepsilon$, the diffusivity becomes much more sensitive to the value of $\gamma$, as indicated in Figure \ref{fig:diff_e1}.




Figure \ref{fig:diffloglog} has been included to illustrate two important properties. Firstly, as the sampling interval goes to zero, the estimator of the diffusivity constant shows linear behavior. This is in agreement with known results for the GBK thermostat\citep{Frank:11} and is an improvement on Langevin thermostats, which would erroneously tend to a constant value as the sampling interval is decreased. Secondly, for large sampling interval the estimator shows an inverse linear tendency. This corresponds simply to the decorrelation of the vortex dynamics.



\subsection{Adaptive determination of multipliers}
Consider the same reduced point vortex model of 8 vortices with $\Gamma = \pm 1$ and assume observations on the energy and momentum are known from a simulation of the full system including the thermal bath. We start such a simulation with an ensemble of $P=100$ initial conditions drawn from the uniform prior. The time step is chosen as $1 \times 10^{-2}$ and the method described in Section \ref{sec:adaptive} for updating the Lagrange multipliers is applied every time unit, i.e. $M=100$. Between subsequent updates of the multipliers, the maximum difference is limited by $|\Delta \lambda_k| \leq 0.1$. When using equilibrium statistics, this limit only affects the beginning of the simulation, when the small sample size used leads to a large variance in the estimators.

\begin{figure}[p]
  \begin{center}
    \includegraphics[width=.33\textwidth]{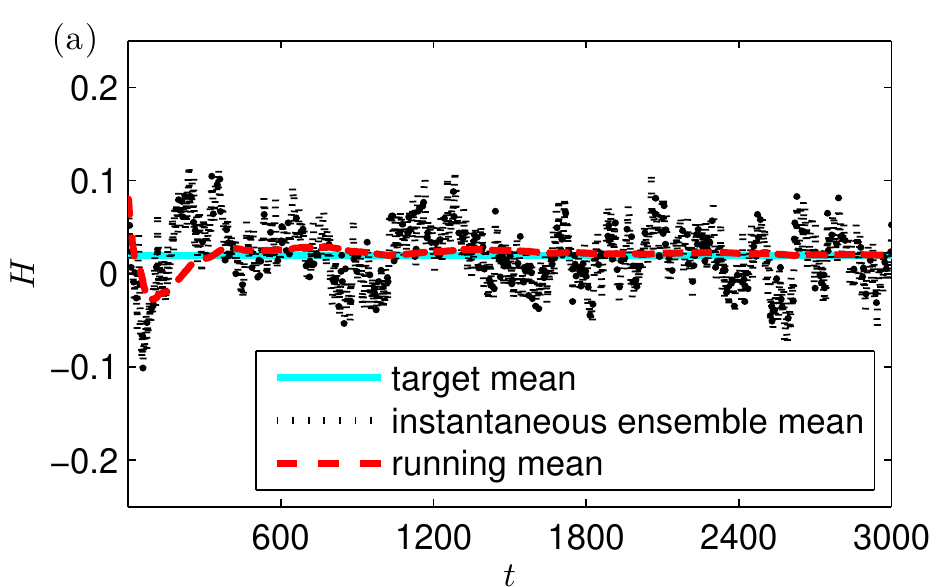}\hfill
    \includegraphics[width=.33\textwidth]{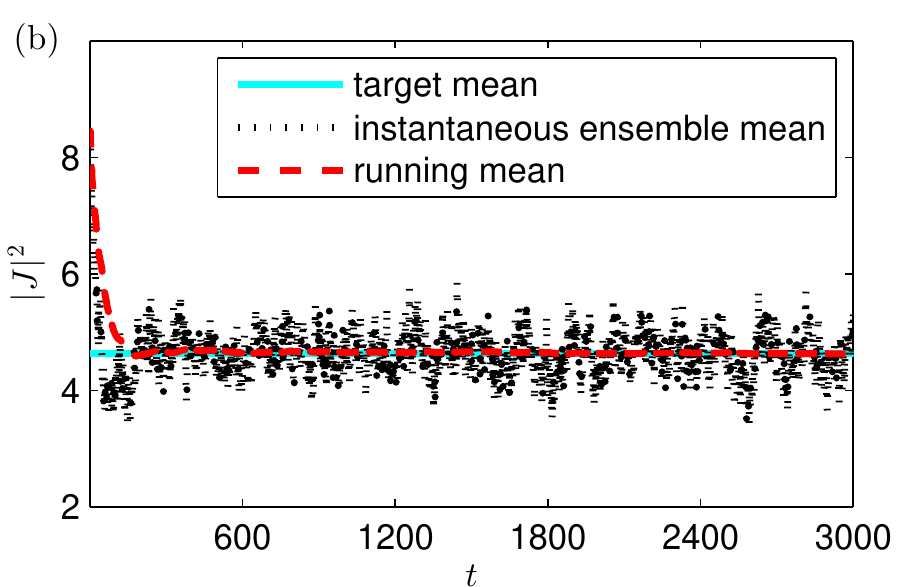}\hfill
    \includegraphics[width=.33\textwidth]{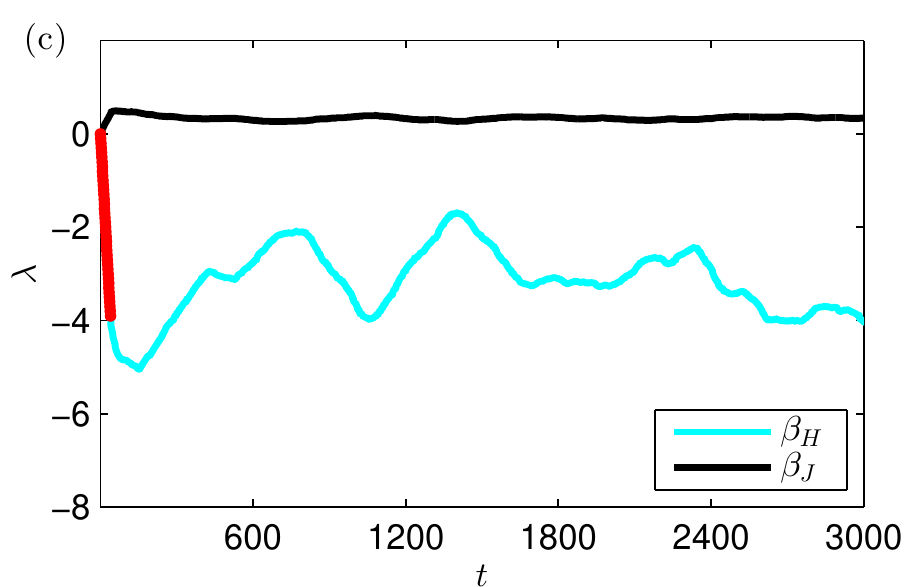}
  \end{center}%
    \caption{Results when using \emph{long-time mean} observations as a target while adaptively determing the Lagrange multipliers. Target observations for Hamiltonian (a) and momentum magnitude (b) are overlaid with the instantaneous ensemble mean (black dotted) and the running ensemble mean (red solid) from simulation. (c): Lagrange multipliers, the red dots indicate time steps at which their rate of change was limited.}
  \label{fig:otf1}
\end{figure}

\begin{figure}[p]
  \begin{center}
    \includegraphics[width=.33\textwidth]{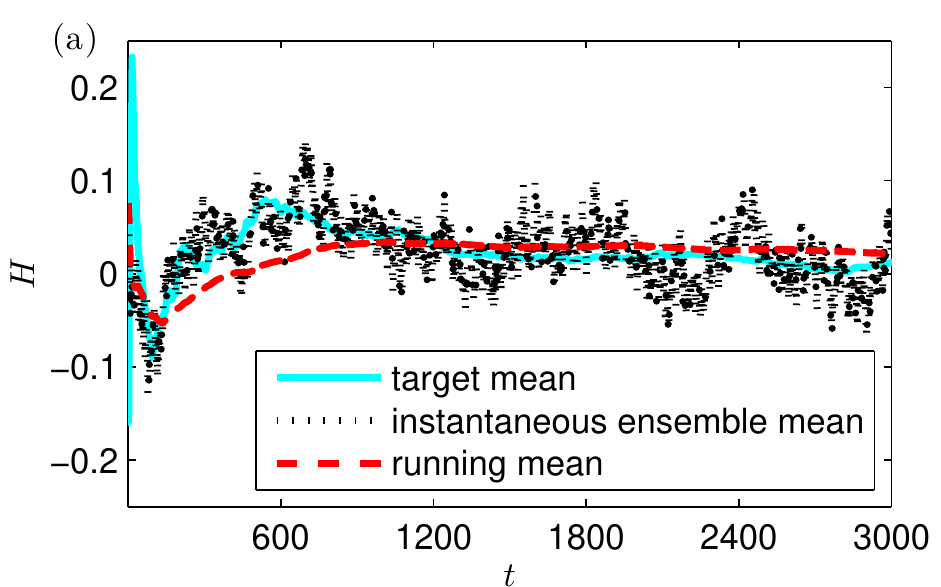}\hfill
    \includegraphics[width=.33\textwidth]{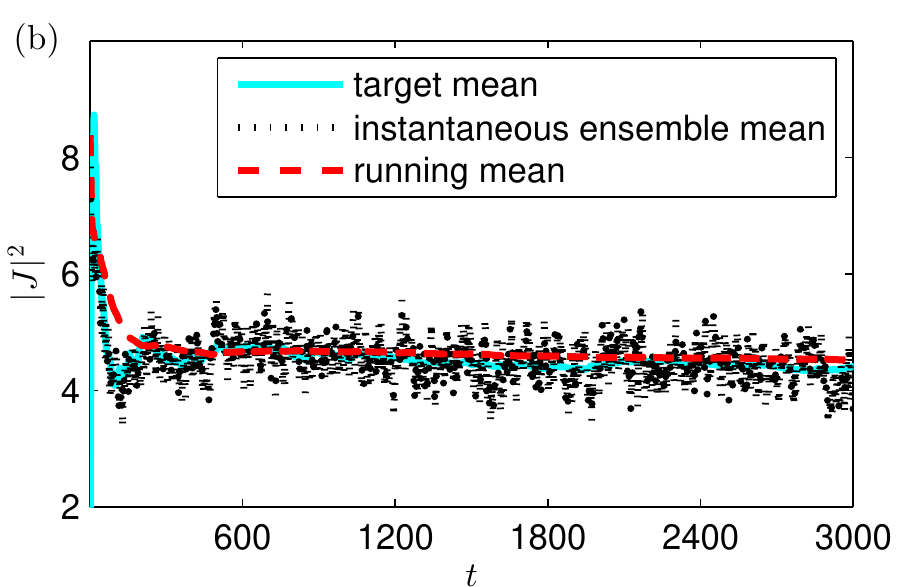}\hfill
    \includegraphics[width=.33\textwidth]{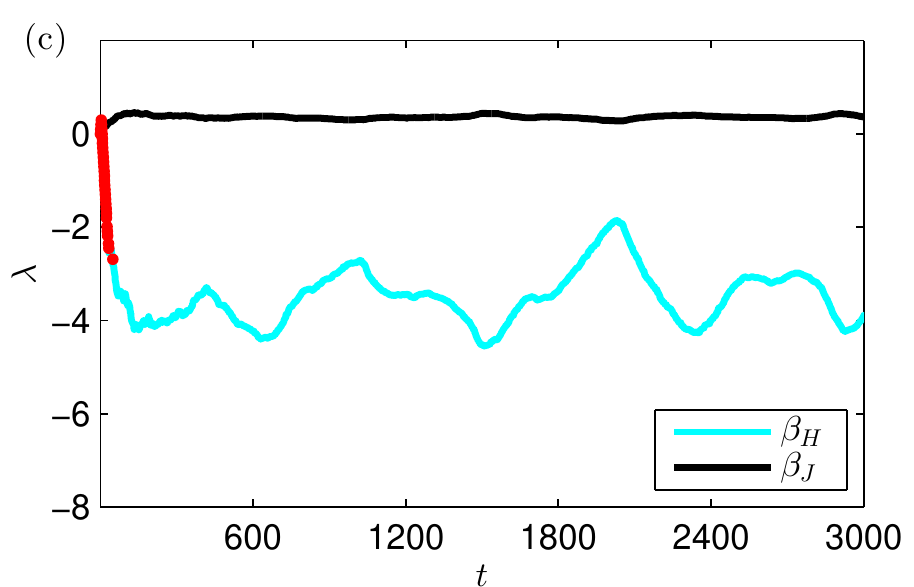}
  \end{center}%
    \caption{Results when using \emph{running mean} observations as a target while adaptively determing the Lagrange multipliers. Target observations for Hamiltonian (a) and momentum magnitude (b) are overlaid with the instantaneous ensemble mean (black dotted) and the running ensemble mean (red solid) from simulation. (c): Lagrange multipliers, the red dots indicate time steps at which their rate of change was limited.}
  \label{fig:otf2}
\end{figure}

\begin{figure}[p]
  \begin{center}
    \includegraphics[width=.33\textwidth]{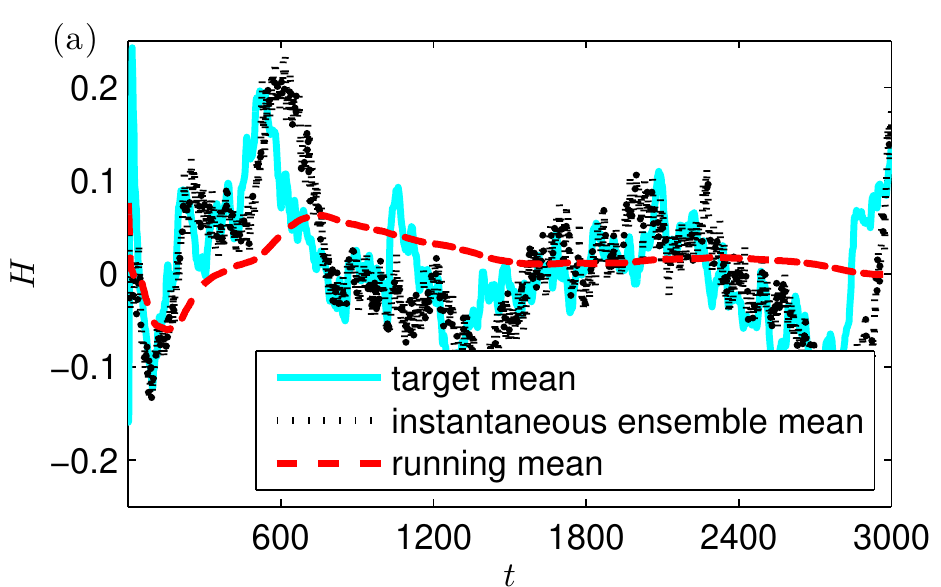}\hfill
    \includegraphics[width=.33\textwidth]{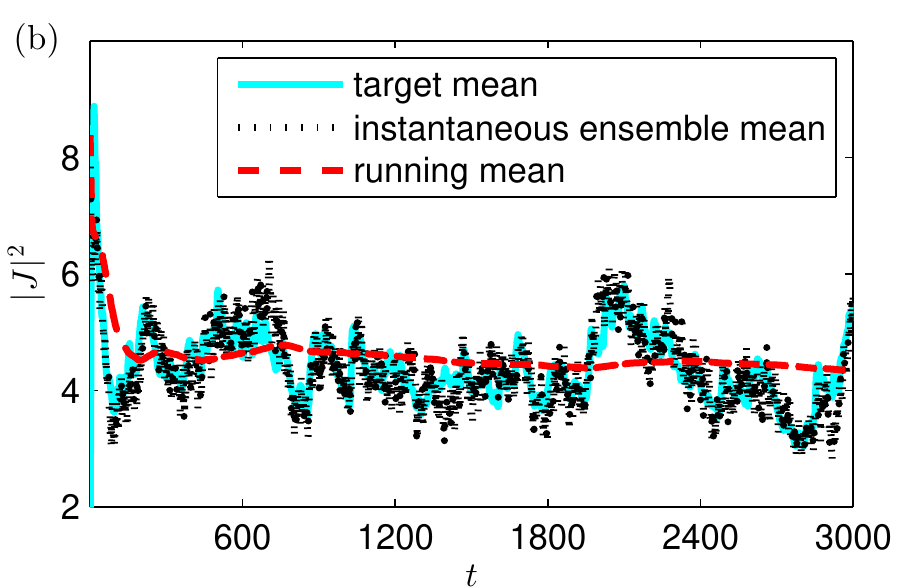}\hfill
    \includegraphics[width=.33\textwidth]{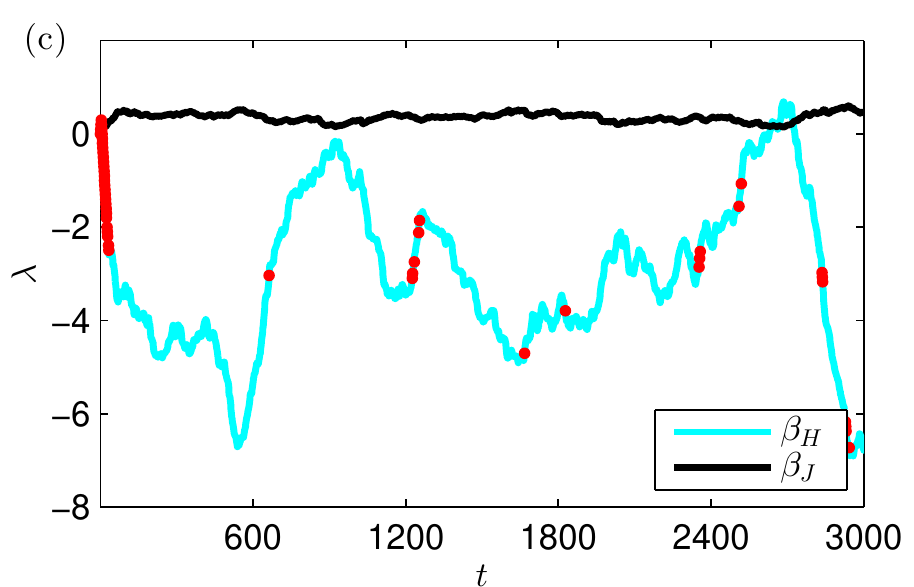}
  \end{center}%
    \caption{Results when using \emph{time-local averaged} observations as a target while adaptively determing the Lagrange multipliers. Target observations for Hamiltonian (a) and momentum magnitude (b) are overlaid with the instantaneous ensemble mean (black dotted) and the running ensemble mean (red solid) from simulation. (c): Lagrange multipliers, the red dots indicate time steps at which their rate of change was limited.}
  \label{fig:otf4}
\end{figure}

The target observation values are taken from a simulation of strong vortices interacting with a thermal bath of weak vortices. Three different averages are used.
\begin{enumerate}
 \item In Figure \ref{fig:otf1} the long time mean is taken and used throughout.
 \item In Figure \ref{fig:otf2} the running mean is used. This reflects the situation where we have no a priori knowledge of the observations, and are continuously feeding new real-time data into the simulation.
 \item In Figure \ref{fig:otf4} a time-localized average of the observable is used. The averaging has a time-scale of a 100 time units. This also corresponds to feeding the simulation new data, but now the assumption of equilibrium is relaxed.
\end{enumerate}


When using either a long time mean observation or a running mean observation, the simulation results tends towards the correct long-time averages. When using time-local averages the simulation averages appear to tend towards a similar value. In all three cases the instantaneous ensemble mean remains close to the (moving) target for both energy and momentum. This is especially notable for the third case, where the target varies over time, but the simulation ensemble mean follows closely, with only a little lag.

The inaccuracies incurred during the first approximately 100 time units indicate that the prior does not match the observed state well. This results in the (negative) growth of $\beta_H$ being limited briefly at the beginning of each simulation. Subsequently, both Lagrange multipliers appear to oscillate irregularly about some mean value for the first two cases. In the case of a shifting target, the Lagrange multipliers vary in time more erratically, as is to be expected. This results in the limiter being active for a few brief periods of the simulation.

\section{Conclusion}
In this article we propose a method for perturbing trajectories of numerical simulations to correct for equilibrium observations.  We treat the restricted case of a Hamiltonian ODE with observations on the set of first integrals of the system.  The approach entails perturbing the solutions using a stochastic thermostat such that they become ergodic in a prescribed invariant measure: the least-biased density corresponding to a maximum entropy treatment of the observations.  

We apply the approach to the case of model reduction in a heterogeneous system of weak and strong point vortices on a sphere, in which observations of the energy $H$ and angular momentum magnitude $|\bs{J}|$ are made on a subsystem consisting of the strong vortices.  A reduced model is constructed by neglecting the weak vortices, and the expectations of the reduced model are corrected using the proposed methodology.  

Numerical experiments confirm that the distributions of the observed quantities $H$ and $|\bs{J}|$ can be well represented using the thermostat technique.  Other equilibrium metrics such as the distribution of distances between like- and opposite-signed vortices are also in agreement across a range of total energy values of the full system, although some discrepancies occur at large positive energies.  

We also investigated the degree to which correction of trajectories for expectations may affect dynamical information in the form of autocorrelation functions and diffusivity.  By decreasing the perturbation parameter $\varepsilon$ of the thermostat, the autocorrelation functions of the unperturbed, reduced system may be precisely recovered.  As $\varepsilon$ is increased, one may increase the diffusivity to values that agree with the full system.  This is consistent with results reported in \cite{Frank:11} in the context of molecular dynamics where it was shown that the GBK thermostat used here approaches Langevin dynamics in the limit of large stochastic forcing.

\section{Acknowledgements}
The work of the first author was supported by a grant financed by the Netherlands Organisation for Scientific Research (NWO), as part of research programme 613.001.009. The third author was supported by grant EP/G036136/1 from the Engineering and Physical Sciences Research Council (UK). 

\appendix

\section{Integration of the two-vortex system\label{sec:2vortex}}
Each pair flow $\phi^{i,j}_{\Delta t}$ is the flow associated with the two-vortex problem $\dot{y} = B \nabla H_{ij}$ for vortex pair $(i,j)$ given by
\begin{align}
  \dot{\bs{x}}_i = \frac{-\Gamma_j}{4\pi} \cdot \frac{\bs{x}_j \times \bs{x}_i}{1-\bs{x}_i \cdot \bs{x}_j}, 
  \label{eq:pvs2a} 
\\
  \dot{\bs{x}}_j = \frac{-\Gamma_i}{4\pi} \cdot \frac{\bs{x}_i \times \bs{x}_j}{1-\bs{x}_i \cdot \bs{x}_j}.
  \label{eq:pvs2b}
\end{align}
This is again a Poisson system, the particular case of \eqref{eq:pvs} with $M=2$. It has a Hamiltonian $H_{ij} = \frac{\Gamma_i \Gamma_j}{8 \pi} \ln\left( 2 - 2\bs{x}_i \cdot \bs{x}_j \right)$ and angular momentum $\bs{J}_{ij} = \Gamma_i \bs{x}_i + \Gamma_j \bs{x}_j$.

Observe that due to the conservation of $H_{ij}$ the denominators in \eqref{eq:pvs2a}--\eqref{eq:pvs2b} are constant. In the numerators of \eqref{eq:pvs2a}--\eqref{eq:pvs2b} we may subsitute $\Gamma_j \bs{x}_j = \bs{J}_{ij} - \Gamma_i \bs{x}_i$ 
and use $\bs{x}_i \times \bs{x}_i = 0$ to find 
\begin{align}
  \dot{\bs{x}}_i = \frac{-1}{4\pi} \cdot \frac{\bs{J}_{ij}}{1-\bs{x}_i \cdot \bs{x}_j} \times \bs{x}_i = \widehat{\bs{a}}\bs{x}_i,
  \label{eq:pvs3a} 
\\
  \dot{\bs{x}}_j = \frac{-1}{4\pi} \cdot \frac{\bs{J}_{ij}}{1-\bs{x}_i \cdot \bs{x}_j} \times \bs{x}_j = \widehat{\bs{a}}\bs{x}_j,
  \label{eq:pvs3b} 
\end{align}
where $\bs{a} = \frac{-1}{4\pi}\frac{\bs{J}}{1-\bs{x}_i \cdot \bs{x}_j}$ and $\widehat{\bs{a}}$ denotes the skew-symmetric matrix associated to the cross product with a vector $\bs{a} = (a_1, a_2, a_3)^T$ by
\[
	\widehat{\bs{a}} = \begin{bmatrix} 0 & -a_3 & a_2 \\ a_3 & 0 & -a_1 \\ -a_2 & a_1 & 0 \end{bmatrix}.
\]
The structure of $\widehat{\bs{a}}$ admits efficient evaluation of the solution $\bs{x}(t) = \exp(\widehat{\bs{a}}t) \bs{x}(0)$ using Rodigues's formula (see, e.g.~\citep{Leimkuhler:04})
\begin{equation*}
  \exp(\widehat{\bs{a}}t) = I + \frac{\sin(\alpha t)}{\alpha} \widehat{\bs{a}} + \frac{1 - \cos(\alpha t)}{\alpha^2} \widehat{\bs{a}}^2,
  \label{eq:rodrigues}
\end{equation*}
where $\alpha = | \bs{a} |$. This is a computationally efficient expression for the given form of $\widehat{\bs{a}}$ because the full matrix exponential need not be computed. Instead we introduce $\tilde{\bs{a}} = \bs{a}/\alpha$ and write
\begin{equation*}
  \exp(\widehat{\bs{a}}t)\bs{x}_0 = \bs{x}_0 + \sin(\alpha t) \tilde{\bs{a}} \times \bs{x}_0 + \left(1 - \cos(\alpha t)\right) \left( \tilde{\bs{a}} \left(\tilde{\bs{a}} \cdot \bs{x}_0\right) - \bs{x}_0 \right).
  \label{eq:rodrigues2}
\end{equation*}

\section{Time integration of the thermostated system}
\label{sec:titd}
In this section we detail the method of integration for the perturbation dynamics and thermostat variable. 

Just as the unperturbed dynamics, the modified double bracket thermostat is a combination of pairwise vortex interactions. Hence it is natural to apply the same splitting to $g$ as was applied to $f$. In fact, the pairwise thermostat interaction can be executed along with the unperturbed dynamics. Recall from Section \ref{sec:ti} that we denote by $\phi^{i,j}_{\Delta t}$ the time $\Delta t$ flow map associated with $f_{ij}$. Similarly we define by $\gamma^{i,j}_{\Delta t}$ the time $\Delta t$ flow map associated with the perturbation dynamics of a vortex pair $(i,j)$. The flow map of the dynamics of the thermostat variable $\xi$ is represented by $\chi_{\Delta t}$. A symmetric composition of these flows is given by
\begin{equation*}
  \Phi_{\Delta t} = 
  \prod_{(i,j) \in C} \left( \phi^{i,j}_{\Delta t/2} \circ \gamma^{i,j}_{\Delta t/2} \right)
\circ 
  \chi_{\Delta t}
\circ 
  \prod_{(i,j) \in C^*} \left( \gamma^{i,j}_{\Delta t/2} \circ \phi^{i,j}_{\Delta t/2} \right).
  \label{eq:psplit}
\end{equation*}
While advancing $\xi$ by $\chi_{\Delta t}$, it is assumed that $y$ is fixed and therefore $h(y)$ is a constant. This means the dynamics of $\xi(t)$ is just an Ornstein-Uhlenbeck process with mean $\frac{h}{\gamma}$ and unit variance. The exact solution is given by
\begin{equation*}
  \chi_{\Delta t} \, \xi_0=\xi_0 e^{-\gamma \Delta t} + \frac{h}{\gamma} (1-e^{-\gamma \Delta t}) + e^{-\gamma \Delta t} W(e^{2\gamma \Delta t}-1).
\end{equation*}

The integration of the perturbed dynamics for a vortex pair $(i,j)$ means integrating the system of ordinary differential equations given by
\begin{align*}
  \dot{\bs{x}}_i &= \frac{\Gamma_j}{4\pi} \bs{x}_i \times \bs{x}_i \times \bs{x}_j
= \frac{\Gamma_i+\Gamma_j}{4\pi} \bs{x}_i \times \bs{x}_i \times \bs{x}_j 
+ \frac{\Gamma_i-\Gamma_j}{4\pi} \bs{x}_i \times \bs{x}_i \times \bs{x}_j,
\\
  \dot{\bs{x}}_j &= \frac{\Gamma_i}{4\pi} \bs{x}_j \times \bs{x}_j \times \bs{x}_i 
= \frac{\Gamma_i+\Gamma_j}{4\pi} \bs{x}_j \times \bs{x}_j \times \bs{x}_i 
- \frac{\Gamma_i-\Gamma_j}{4\pi} \bs{x}_j \times \bs{x}_j \times \bs{x}_i.
\end{align*}
The equations have been split into a symmetric and an anti-symmetric part. A symmetric composition may once more be applied to integrate the two parts. In our case however, the thermostated system will consist only of vortices with strength $\pm\Gamma_\text{strong}$ and thus each vortex pair interaction is either fully symmetric or fully antisymmetric.

By introducing $\alpha_S = \frac{\Gamma_i+\Gamma_j}{4\pi}$ the symmetric part may be written as
\begin{align}
  \dot{\bs{x}}_i &= \alpha_S \bs{x}_i \times \bs{x}_i \times \bs{x}_j = \alpha_S \left( \bs{x}_j \times \bs{x}_i \right) \times \bs{x}_i,
  \label{eq:ppvssa} 
\\
  \dot{\bs{x}}_j &= \alpha_S \bs{x}_j \times \bs{x}_j \times \bs{x}_i =-\alpha_S \left( \bs{x}_j \times \bs{x}_i \right) \times \bs{x}_j.
  \label{eq:ppvssb}
\end{align}
These dynamics are symmetric with respect to the plane equidistant between the two vortices. This is because the dynamics are rotations in opposite direction about the same vector $\bs{x}_j \times \bs{x}_i$ and this vector must lie in said plane. Furthermore the dynamics of \eqref{eq:ppvssa}--\eqref{eq:ppvssb} does not allow the vortex pair to pass through the position where they are antipedian, as in this case their cross product is zero. All in all this means the final position of the two vortices is uniquely determined by their chord distance.

The change in the distance between the two vortices can be represented by the change of their inner product
\begin{align*}
 \pd{}{t}\left( \bs{x}_i \cdot \bs{x}_j \right) 
&= \dot{\bs{x}}_i \cdot \bs{x}_j + \bs{x}_i \cdot \dot{\bs{x}}_j \\
&= 2\alpha_S \left( (\bs{x}_i \cdot \bs{x}_j)^2 - 1 \right) 
\end{align*}
The solution to this differential equation is given by
\begin{equation*}
  \bs{x}_i \cdot \bs{x}_j |_{t=\Delta t} = \tanh \left( 2\alpha \Delta t - \artanh \left( \bs{x}_i \cdot \bs{x}_j \right) |_{t=0} \right).
\end{equation*}

In the anti-symmetric part we introduce $\alpha_A = \frac{\Gamma_i-\Gamma_j}{4\pi}$ to write
\begin{align*}
  \dot{\bs{x}}_i &=  \alpha_A \bs{x}_i \times \bs{x}_i \times \bs{x}_j,
\\
  \dot{\bs{x}}_j &= -\alpha_A \bs{x}_j \times \bs{x}_j \times \bs{x}_i.
\end{align*}
By rearranging the order of the cross product we achieve
\begin{align}
  \dot{\bs{x}}_i &= \alpha_A \left( \bs{x}_j \times \bs{x}_i \right) \times \bs{x}_i = \widehat{a}_A \bs{x}_i,
  \label{eq:ppvsaa1} 
\\
  \dot{\bs{x}}_j &= \alpha_A \left( \bs{x}_j \times \bs{x}_i \right) \times \bs{x}_j = \widehat{a}_A \bs{x}_j.
  \label{eq:ppvsab1}
\end{align}
The vector $\bs{a}_A$ implicity defined by \eqref{eq:ppvsaa1}--\eqref{eq:ppvsab1} can simply be shown to be constant under the antisymmetric flow. This means the dynamics of the anti-symmetric part may be integrated by using Rodigues's formula, similar to the unperturbed dynamics \eqref{eq:pvs3a}--\eqref{eq:pvs3b}.

\section*{References}
\bibliographystyle{elsarticle-num}

\end{document}

%% file: macros.tex
\let\oldsqrt\sqrt
\def\sqrt{\mathpalette\DHLhksqrt}
\def\DHLhksqrt#1#2{%
\setbox0=\hbox{$#1\oldsqrt{#2\,}$}\dimen0=\ht0
\advance\dimen0-0.2\ht0
\setbox2=\hbox{\vrule height\ht0 depth -\dimen0}%
{\box0\lower0.4pt\box2}}
\newcommand{\R}{\mathbf{R}}

\newcommand{\artanh}{\mathop{\mathrm{artanh}}\nolimits} 
\renewcommand{\epsilon}{\varepsilon}

\makeatletter
\def\psls@dotteddashed{%
[ 0 \psk@dotsep CLW add \psk@dash ] 0 setdash  1 setlinecap  stroke}
\makeatother

\newcommand{\bs}[1]{\boldsymbol{#1}}

\newcommand{\pd}[2]{\frac{\partial #1}{\partial #2}}

\newcommand{\poibra}[2]{\left\lbrace #1 , #2 \right\rbrace}
\newcommand{\ensbra}[1]{\left\langle #1 \right\rangle}


%% file: tabHs_Lagrange.tex
\begin{tabular}{l|r|r|r|r|r}
&\textbf{$\beta_H$}&\textbf{$\beta_J$}&\textbf{$\gamma_H$}&\textbf{$\gamma_J$}&\textbf{$\gamma_{HJ}$}\\\hline
\textbf{$H_\text{full} = -2$}&$5.98$&$-0.20$&$0.69$&$0.41 \times 10^{-3}$&$-0.04$\\
\textbf{$H_\text{full} = -1$}&$2.89$&$-0.03$&$2.67$&$9.77 \times 10^{-3}$&$-0.33$\\
\textbf{$H_\text{full} = 0$}&$-0.76$&$0.20$&$3.38$&$9.97 \times 10^{-3}$&$-0.37$\\
\textbf{$H_\text{full} = 1$}&$-3.54$&$0.37$&$4.29$&$15.31 \times 10^{-3}$&$-0.54$\\
\textbf{$H_\text{full} = 2$}&$-6.42$&$0.53$&$4.45$&$14.05 \times 10^{-3}$&$-0.51$\\
\end{tabular}

%% file: tabNs_Lagrange.tex
\begin{tabular}{l|r|r|r|r|r}
&\textbf{$\beta_H$}&\textbf{$\beta_J$}&\textbf{$\gamma_H$}&\textbf{$\gamma_J$}&\textbf{$\gamma_{HJ}$}\\\hline
\textbf{$M_\text{full} = 36$}&$-1.51$&$1.35$&$27.75$&$117.15 \times 10^{-3}$&$-3.07$\\
\textbf{$M_\text{full} = 72$}&$-4.27$&$0.82$&$8.71$&$37.79 \times 10^{-3}$&$-1.12$\\
\textbf{$M_\text{full} = 144$}&$-0.97$&$0.32$&$6.70$&$25.80 \times 10^{-3}$&$-0.82$\\
\textbf{$M_\text{full} = 288$}&$-0.76$&$0.20$&$3.38$&$9.97 \times 10^{-3}$&$-0.37$\\
\textbf{$M_\text{full} = 576$}&$-1.09$&$0.13$&$0.87$&$3.08 \times 10^{-3}$&$-0.10$\\
\end{tabular}